\documentstyle[psfig]{mn}

\newif\ifAMStwofonts

\ifoldfss
  \ifCUPmtlplainloaded \else
    \NewTextAlphabet{textbfit} {cmbxti10} {}
    \NewTextAlphabet{textbfss} {cmssbx10} {}
    \NewMathAlphabet{mathbfit} {cmbxti10} {} 
    \NewMathAlphabet{mathbfss} {cmssbx10} {} 
  \fi
  \ifAMStwofonts
    \ifCUPmtlplainloaded \else
      \NewSymbolFont{upmath} {eurm10}
      \NewSymbolFont{AMSa} {msam10}
      \NewMathSymbol{\upi}     {0}{upmath}{19}
      \NewMathSymbol{\umu}     {0}{upmath}{16}
      \NewMathSymbol{\upartial}{0}{upmath}{40}
      \NewMathSymbol{\leqslant}{3}{AMSa}{36}
      \NewMathSymbol{\geqslant}{3}{AMSa}{3E}

       \let\le=\leqslant
      \let\geq=\geqslant \let\ge=\geqslant
    \fi
  \fi
\fi 

\ifnfssone
  \newmathalphabet{\mathit}
  \addtoversion{normal}{\mathit}{cmr}{m}{it}
  \addtoversion{bold}{\mathit}{cmr}{bx}{it}
  \newmathalphabet{\mathbfit} 
  \addtoversion{normal}{\mathbfit}{cmr}{bx}{it}
  \addtoversion{bold}{\mathbfit}{cmr}{bx}{it}
  \newmathalphabet{\mathbfss} 
  \addtoversion{normal}{\mathbfss}{cmss}{bx}{n}
  \addtoversion{bold}{\mathbfss}{cmss}{bx}{n}
  \ifAMStwofonts
    \ifCUPmtlplainloaded \else
      \UseAMStwoboldmath
      \makeatletter
      \new@mathgroup\upmath@group
      \define@mathgroup\mv@normal\upmath@group{eur}{m}{n}
      \define@mathgroup\mv@bold\upmath@group{eur}{b}{n}
      \edef\UPM{\hexnumber\upmath@group}
      \new@mathgroup\amsa@group
      \define@mathgroup\mv@normal\amsa@group{msa}{m}{n}
      \define@mathgroup\mv@bold\amsa@group{msa}{m}{n}
      \edef\AMSa{\hexnumber\amsa@group}
      \makeatother
      \mathchardef\upi="0\UPM19
      \mathchardef\umu="0\UPM16
      \mathchardef\upartial="0\UPM40
      \mathchardef\leqslant="3\AMSa36
      \mathchardef\geqslant="3\AMSa3E

       \let\le=\leqslant
      \let\geq=\geqslant \let\ge=\geqslant
    \fi
  \fi
\fi 

\ifnfsstwo
  \DeclareMathAlphabet{\mathbfit}{OT1}{cmr}{bx}{it}
  \SetMathAlphabet\mathbfit{bold}{OT1}{cmr}{bx}{it}
  \DeclareMathAlphabet{\mathbfss}{OT1}{cmss}{bx}{n}
  \SetMathAlphabet\mathbfss{bold}{OT1}{cmss}{bx}{n}
  \ifAMStwofonts
    \ifCUPmtlplainloaded \else
      \DeclareSymbolFont{UPM}{U}{eur}{m}{n}
      \SetSymbolFont{UPM}{bold}{U}{eur}{b}{n}
      \DeclareSymbolFont{AMSa}{U}{msa}{m}{n}
      \DeclareMathSymbol{\upi}{0}{UPM}{"19}
      \DeclareMathSymbol{\umu}{0}{UPM}{"16}
      \DeclareMathSymbol{\upartial}{0}{UPM}{"40}
      \DeclareMathSymbol{\leqslant}{3}{AMSa}{"36}
      \DeclareMathSymbol{\geqslant}{3}{AMSa}{"3E}

       \let\le=\leqslant
      \let\geq=\geqslant \let\ge=\geqslant
    \fi
  \fi
\fi 

\ifCUPmtlplainloaded \else
  \ifAMStwofonts \else 
    \def\upi{\pi}
    \def\umu{\mu}
    \def\upartial{\partial}
  \fi
\fi

\title{Globular cluster formation from
gravitational tidal effects  of merging and interacting
galaxies}
\author[K. Bekki,  Duncan  A. Forbes, M. A. Beasley,  W.  J. Couch]
       {K. Bekki,${}^1$   Duncan A. Forbes${}^2$, M. A. Beasley${}^2$,  and   W. J. Couch${}^1$\\
        ${}^1$School of Physics, University of New South Wales, Sydney 2052, NSW, Australia \\
        ${}^2$Centre for Astrophysics \& Supercomputing, Swinburne University
of Technology, Hawthorn, VIC,  3122
Australia}
\date{Accepted 
      Received
      in original form 2001}

\pagerange{\pageref{firstpage}--\pageref{lastpage}}
\pubyear{1994}

\begin{document}

\maketitle

\label{firstpage}

\begin{abstract}

We investigate the spatial, kinematic and chemical properties of
globular cluster systems formed in merging and interacting
galaxies using N-body/SPH simulations. Although we can not
resolve individual clusters in our simulation, we assume that 
they form in collapsing molecular clouds when the local external
gas pressure exceeds 10$^5$ $k_B$ (where $k_B$ is the Boltzmann
constant). Several simulations are carried out for a range of
initial conditions and galaxy mass ratios. The input model spirals
are given a halo globular cluster system similar to those
observed for the Milky Way and M31.
Gravitational tidal effects during galaxy 
merging and interaction leads to a dramatic increase in gas
pressure, which exceeds our threshold and hence triggers 
new globular cluster formation. 
We investigate the properties of the globular cluster system in the
remnant galaxy, such as number density, specific frequency,
kinematic properties and metallicity distribution. Different
orbital conditions and mass ratios give rise to a range in
globular cluster properties, particularly for the interaction
models. Our key results are the following:
The newly formed metal-rich clusters are concentrated at the
centre of the merger remnant elliptical, whereas the metal-poor
ones are distributed to the outer parts due to strong angular
momentum transfer. 
The dissipative merging of {\it present day} spirals, including
chemical evolution, results in
metal-rich clusters with a mean metallicity that is super-solar,
i.e. much higher than is observed in elliptical galaxies. 
If elliptical galaxies form by dissipative major mergers, then
they must do so at very early epochs when their discs contained low
metallicity gas. 
Our simulations show that the specific frequency can be increased
in a dissipative major merger. However, when this occurs it
results in a ratio of metal-poor to metal-rich clusters is less than
one, contrary to the ratio observed in many elliptical galaxies. 

\end{abstract}


\section{Introduction}

There have been many previous models and 
scenarios for the formation of globular
clusters and their systems 
(e.g., Peebles \& Dicke 1968; Searle \& Zinn 1978;
Fall \& Rees 1985; Fall \& Rees 1988;
Zinnecker, Keable, \& Dunlop 1988;
Larson 1987, 1988;
Kang et al. 1990; 
Ashman \& Zepf 1992;
Freeman 1993;
Harris \& Pudritz 1994; Elmegreen \& Efremov 1997; Forbes et al. 1997;
McLaughlin 1999;
Ashman \& Zepf 2001; Bekki \& Couch 2001; Cen 2001; Weil \& Pudritz 2001;
Beasley et al. 2002 and Bekki \& Chiba 2002: 
See Harris 1991 and Ashman \& Zepf 1998 for a review).
These studies have contributed
to a  better understanding of the physical 
conditions of globular cluster formation at low and high
redshifts, in addition to providing insight into the
the physical origins of the observed scaling relations 
of globular cluster systems.

However, they have not addressed  
the observed structural, kinematical, and chemical
properties of globular cluster systems   
{\it in a self-consistent manner} through the detailed modelling
of collapsing  molecular clouds. 
Furthermore, given the observational fact that young super star clusters
are often located in star-burst regions 
(e.g., Ashman \& Zepf 1998; 2001),
it is critically important 
to investigate physical properties of globular clusters
formed by the induced collapse of molecular clouds 
{\it for the case of galaxy merging and interaction}.
Comparing the predicted 
properties of globular cluster systems
with the corresponding observational ones in E/S0 galaxies
is important,
because recent  observations
provide a  rich data set of metallicity and kinematics
for  globular  clusters around giant elliptical galaxies 
(e.g., Kissler-Patig et al. 1999; Cohen et al. 1998
Beasley et al. 2000; Bridges 2001; Forbes 2001; Forbes et al.  2001).
Furthermore  better understanding   of
the nature  of globular cluster formation 
in {\it  present-day}
merging and interacting galaxies
can provide clues to their formation 
{\it at  high redshift}, where merging and interaction 
 was  much more frequent.

The purpose of this paper is to numerically 
investigate the kinematical and chemical properties of 
globular clusters formed in merging and interacting spiral galaxies.
We adopt the plausible assumption that the high pressure of warm
interstellar gas ($P_{\rm gas}$ $>$ $10^5$ $k_{\rm B}$: $k_{\rm
B}$ is Boltzmann's constant) can induce the global collapse of
giant molecular clouds to form massive compact star clusters
corresponding to super star clusters or progenitor objects of
globular clusters (Jog \& Solomon 1992; Elmegreen \& Efremov
1997).  We mainly investigate the following five points: (1) why
is globular cluster formation more efficient in merging galaxies
than in isolated spiral galaxies, (2) how does the specific frequency
($S_{\rm N}$) of globular cluster systems change during merging, (3)
what are the fundamental properties (e.g., structure, kinematics, and
metallicity distribution) of newly formed globular
cluster systems, (4) how do the physical properties of globular clusters
formed during merging depend on orbital configuration, mass-ratio
of the merging spirals, gas mass fraction, and merging epoch, and
(5) are there any differences in the details of globular
cluster formation in tidally interacting galaxies versus merging
galaxies.
We also stress that better understanding the effects of galactic tides 
on the dynamical and chemical evolution
of interstellar medium is of  primary importance  for 
clarifying  the unresolved problems related to globular cluster
formation in the low and high redshift universe. 

The plan of the paper is as follows: In the next section,
we describe our  numerical model for  globular cluster formation
based on the adopted molecular cloud collapse scenario.   
In \S 3, we 
present the numerical results
on structural, kinematical, and chemical properties
of globular clusters formed in merging and interacting  galaxies.
In \S 4, we  compare the basic properties of globular cluster systems
from our model predictions
with the observations. We also present additional predictions which can be tested
against future observations.
We summarise our  conclusions in \S 5.

\section{The model}

\subsection{Disk Model}

Since our numerical methods for modelling chemodynamical
evolution of gas-rich galaxies and the details of the adopted
TREESPH codes have already been described by Bekki \& Shioya
(1998) and by Bekki (1995), respectively, we give only a brief
review here.  We construct models of gas-rich and star-forming
spiral galaxies by using the Fall-Efstathiou model (1980).  The
total mass and the size of a spiral in the standard disc model
are $M_{\rm d}$ and $R_{\rm d}$, respectively. In this standard
disc model, the disc evolves without any merging and tidal
interaction with other galaxies (This is an isolated model that
will be compared with merger or interaction models described
later).  Henceforth, all masses and lengths are measured in units
of $M_{\rm d}$ and $R_{\rm d}$, respectively, unless
specified. Velocity and time are measured in units of $v$ = $
(GM_{\rm d}/R_{\rm d})^{1/2}$ and $t_{\rm dyn}$ = $(R_{\rm
d}^{3}/GM_{\rm d})^{1/2}$, respectively, where $G$ is the
gravitational constant and assumed to be 1.0 in the present
study. If we adopt $M_{\rm d}$ = 6.0 $\times$ $10^{10}$ $ \rm
M_{\odot}$ and $R_{\rm d}$ = 17.5 kpc as a fiducial value, then
$v$ = 1.21 $\times$ $10^{2}$ km/s and $t_{\rm dyn}$ = 1.41
$\times$ $10^{8}$ yr, respectively. In the present study, the
mass and the size of a disc are assumed to be free parameters
represented by $m_{\rm d}$ and $r_{\rm d}$, respectively, and
$m_{\rm d}$ = 1, $r_{\rm d}$ = 1 for the standard disc model.
The relation between $m_{\rm d}$ and $r_{\rm d}$ is chosen such
that it is consistent with the Freeman's law (1970) for most
models.

In the standard disc model, the rotation curve becomes nearly
flat at $R$ = 0.35 (where $R$ is distance from the centre of the
disc) with the maximum rotational velocity $v_{\rm m}$ = 1.8 in
our units (220 km s$^{-1}$).  The corresponding total mass of the
dark matter halo (within 1.5$R_{\rm d}$) is 4.0 in our units.
The velocity dispersion of halo component at a given point is set
to be isotropic and given according to the virial theorem.  The
radial ($R$) and vertical ($Z$) density profile of the disc are
assumed to be proportional to $\exp (-R/R_{0}) $ with scale
length $R_{0}$ = 0.2 and to ${\rm sech}^2 (Z/Z_{0})$ with scale
length $Z_{0}$ = 0.04 in our units, respectively.  The central
bulge with the mass of 0.25 and the size of 0.2 in our units is
represented by the Plummer model with a scale length of 0.04.
The corresponding bulge-to-disc ratio is 0.25 in the disc model.
The Galaxy and M31 are observed to have 160 $\pm$ 20 and 400
$\pm$ 55 globular clusters, respectively (van den Bergh 1999).
Guided by these observations, the spiral is assumed to have 200
old, metal-poor globular clusters with the number density
distribution the same as that of the Galaxy globular cluster
system (i.e., $\rho (r) \sim r^{-3.5}$). In addition to the
rotational velocity made by the gravitational field of disc,
bulge, and dark halo components, the initial radial and azimuthal
velocity dispersions are given to the disc component according to
the epicyclic theory with Toomre's parameter $Q$ = 1.2.  The
vertical velocity dispersion at given radius is set to be 0.5
times as large as the radial velocity dispersion at that point,
as is consistent with the observed trend of the Milky Way (e.g.,
Wielen 1977).

An isothermal equation of state is used for the gas with
temperatures of $10^4$ K and 2500 K.  We describe mostly the
results of the models with the gaseous temperature $10^4$ K.
Chemical enrichment through star formation (described below) in
merging and interacting galaxies is assumed to proceed both
locally and instantaneously in the present study.  The fraction
of gas returned to the interstellar medium in each stellar
particle and the chemical yield are 0.3 and 0.02, respectively.
The initial metallicity, $Z_{\ast}$, for each gaseous particle at
a given galactic radius $R$ (kpc) from the centre of the disc is
given according to the observed relation $Z_{\ast} = Z_{\rm g}(0)
\times {10}^{-0.197 \times (R/3.5)}$ typical of late-type spiral
galaxies with $Z_{\rm g}(0)$ = 0.06 (e.g., Zaritsky, Kennicutt,
\& Huchra 1994). In the present study, we consider that $Z_{\rm
g}(0)$ is a free parameter which should be changed with redshift
of merging and interacting galaxy: At a higher redshift, $Z_{\rm
g}(0)$ could be lower than the present-day value of typical
spiral galaxies.  The total particle number in each simulation is
23178 for the collisionless components, 20000 for the collisional
ones and the parameter of gravitational softening is set to be
fixed at 0.038 in our units.

We investigate both merger and tidal interaction models.  In the
simulation of a merging or interacting galaxy, the orbit of the
two spirals is set to be initially in the $xy$ plane and the
distance between the centre of mass of the two spirals ($r_{\rm
in}$) is 4$r_{\rm d}$ (70 kpc for the model with $m_{\rm d}$ =
1.0). The pericentre distance ($r_{\rm p}$) and the orbital
eccentricity ($e_{\rm p}$) are assumed to be free parameters
which control orbital angular momentum and energy of the merging
or interacting galaxy.  For most merger models, $r_{\rm p}$ and
$e_{\rm p}$ are set to be 0.5 $r_{\rm d}$ (8.75 kpc for $m_{\rm
d}$ = 1.0) and 1.0, respectively.  For most tidal interaction
models, $r_{\rm p}$ and $e_{\rm p}$ are set to be 2.0 $r_{\rm d}$
and 1, respectively.  The spin of each galaxy in a merging and
interacting galaxy is specified by two angles $\theta_{i}$ and
$\phi_{i}$ (in units of degrees), where the suffix $i$ is used to
identify each galaxy.  Here, $\theta_{i}$ is the angle between
the $z$ axis and the vector of the angular momentum of the disc,
and $\phi_{i}$ is the azimuthal angle measured from $x$ axis to
the projection of the angular momentum vector of the disc onto
the $xy$ plane. We specifically investigate the following five
models with different disc inclinations with respect to the
orbital plane: A prograde-prograde model represented by ``PP''
with $\theta_{1}$ = 0, $\theta_{2}$ = 30, $\phi_{1}$ = 0, and
$\phi_{2}$ = 0, a prograde-retrograde (``PR'') with $\theta_{1}$
= 0, $\theta_{2}$ = 210, $\phi_{1}$ = 0, and $\phi_{2}$ = 0, a
retrograde-retrograde (``RR'') with $\theta_{1}$ = 180,
$\theta_{2}$ = 210, $\phi_{1}$ = 0, and $\phi_{2}$ = 0, a highly
inclined model (``HI'') with $\theta_{1}$ = 60, $\theta_{2}$ =
60, $\phi_{1}$ = 90, and $\phi_{2}$ = 0 and an Antennae model
(``AN'') with $\theta_{1}$ = 60, $\theta_{2}$ = $-60$, $\phi_{1}$
= 210, and $\phi_{2}$ = 210. In the case of major merging, the
``AN'' model with $r_{\rm p}$ = 1.0 and $e_{\rm p}$ = 0.5 shows
morphological properties strikingly similar to those of ``the
Antennae'' NGC 4038/39.  The model with larger orbital angular
momentum ($r_{\rm p}$ = 1.0) and a PP orbital configuration is
labelled as ``LA''

\begin{figure*}
\psfig{file=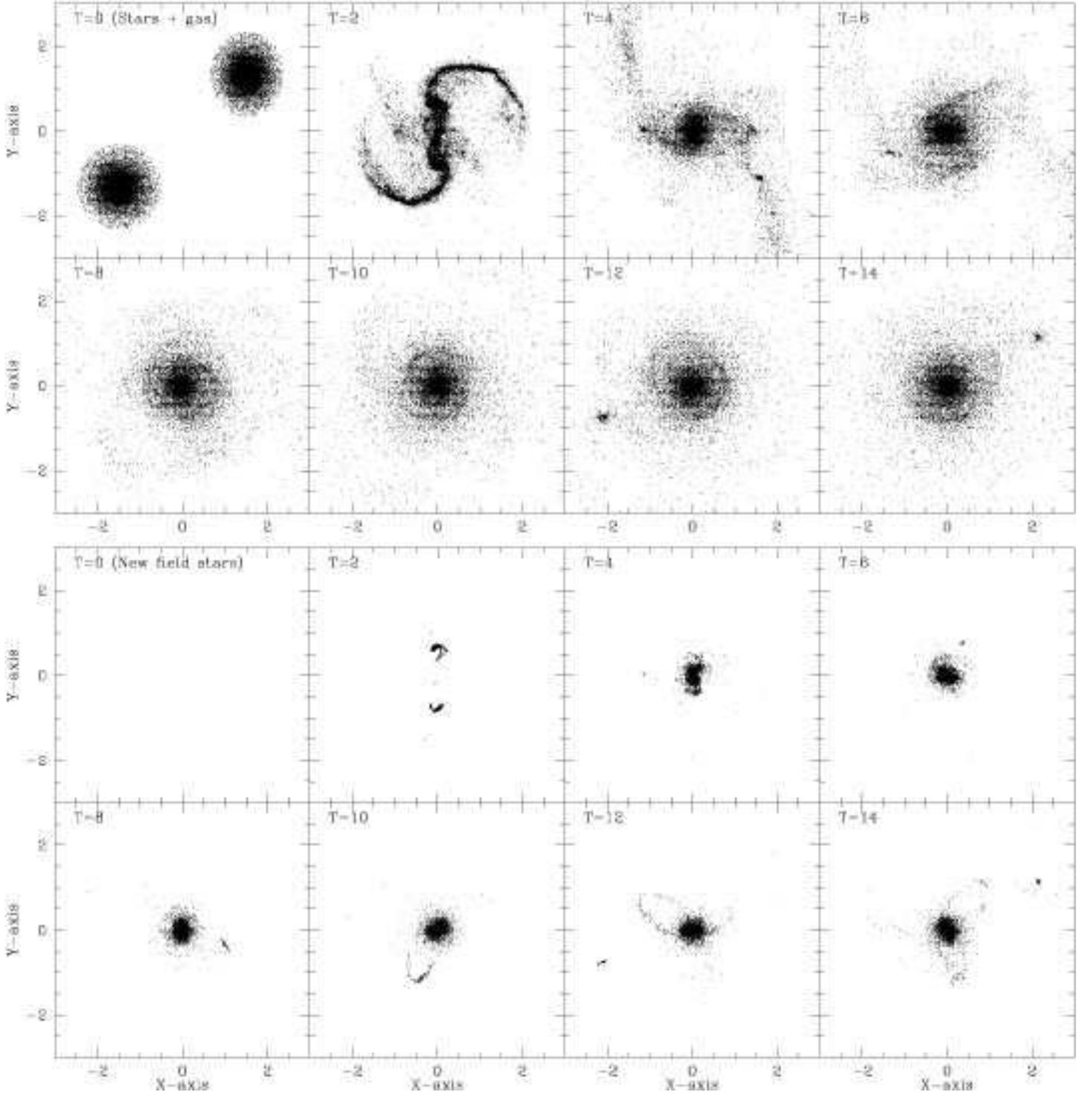}
\caption{ 
Morphological evolution of old stellar and gaseous components (upper eight panels)
and new field stars (lower eight panels) 
in the fiducial merger model M1 projected onto the  $x$-$y$ plane.
The time $T$ (in our units) represents the time that has elapsed since
the simulation starts. 
This figure accordingly describes $\sim$ 2 Gyr dynamical evolution of a nearly prograde-prograde
gas-rich major merger with star formation.
The scale is given in our units, and accordingly each frame measures
105 kpc on a side. 
}
\label{Figure. 1}
\end{figure*}


\begin{figure*}
\psfig{file=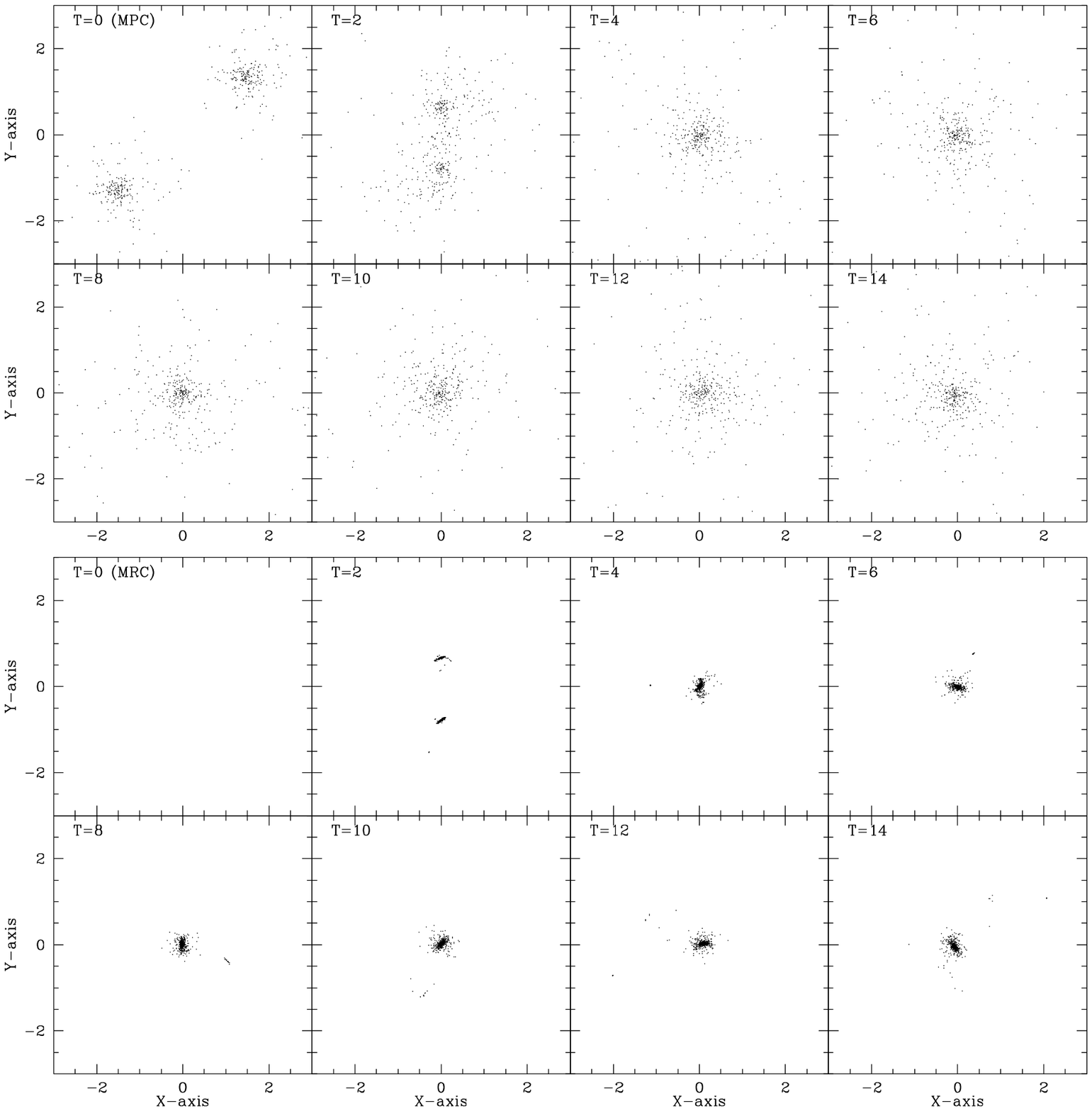}
\caption{ 
The same as Fig. 1 but for metal-poor globular clusters  (MPC)
in the upper eight panels and 
for metal-rich clusters  (MRC) in the  lower eight panels.
}
\label{Figure. 2}
\end{figure*}

\subsection{Formation of field stars and globular clusters}

We adopt the following  star formation laws to
convert gaseous particles into either globular clusters or field stars.
We stress that throughout the  paper,
we distinguish between the  newly formed ``field stars'' and  stars 
initially located within the disc and bulge (referred to as ``old stars''). 
For globular clusters,  we adopt the formation model by
Jog \& Solomon (1992) and Elmegreen \& Efremov (1997),
in which interstellar gaseous pressure ($P_{\rm gas}$) in star forming regions of
a galaxy drives the collapse of pressure confined, 
magnetised self-gravitating   molecular clouds
to  form compact clusters, providing  $P_{\rm gas}$ is larger than the surface
pressure ($P_{\rm s}$) of the clouds: 
\begin{equation}
P_{\rm gas} \ge  P_{\rm s} \sim 2.0\times 10^5 k_{\rm B}. \;
\end{equation}  
Because only a small fraction of gas in a molecular cloud 
can be converted into stars (e.g., Larson 1987),
we introduce the formation probability $C_{\rm gc}$ with which
{\it one}  new cluster  is converted from a gas particle if the gas pressure
is larger than $P_{\rm s}$ = $2.0\times 10^5 k_{\rm B}$.
For most  models, $C_{\rm gc}$ is 0.1, which implies
that a cluster forms with a 10 \% probability  
when the gas pressure is larger than $P_{\rm s}$ = $2.0\times 10^5 k_{\rm B}$.
Since $C_{\rm gc}$ is a largely unconstrained quantity,
we also investigate the models with 
$C_{\rm gc}$ =  0.5 and 1.0.
Although this formation model is apparently different from cloud collision
models by Kumai, Basu, \& Fujimoto (1993) and Fujimoto \& Kumai (1997)
for interacting galaxies,
we do not intend to discuss which model is more plausible
for globular cluster formation here.

For field stars, we adopt the Schmidt law (Schmidt 1959) with an  exponent
of 1.5 (Kennicutt 1989). 
The coefficient of the Schmidt law is chosen such that
the standard isolated disc model without cluster formation
exhibits a   mean field star formation rate of a few ${\rm M}_{\odot}$ ${\rm yr}^{-1}$
(corresponding to typical star formation rate in disc galaxies)
during the 2 Gyr evolution. 
Although this model for star formation is based on
the observed {\it current} star formation law in nearby disc galaxies,
which may not be simply applied to merging and interacting galaxies, 
its details do not affect
the derived results for forming clusters.
Also, although we cannot investigate the detailed physical processes of
cluster formation in the present {\it global}
(from $\sim$ 100 pc to 10 kpc scale) simulation, we expect that the adopted
phenomenological approach enables us to
identify the plausible formation sites of globular clusters.
The newly formed clusters with larger metallicities are referred to as 
metal-rich clusters (`` MRC'')
in order that we can distinguish
them from  metal-poor ones  (``MPC'') initially in the spirals. 
Therefore the remnant of a merging and interacting spiral galaxy
is composed of a dark halo, bulge, old disc  stars, new field stars,
the MPC (with the initial number of 200 in a spiral), and the MRC.

\subsection{Main  points of analysis}

The specific frequency is defined as follows (Harris \& van den Bergh 1981):
\begin{equation}
S_{\rm N}=N_{\rm gc} \times 10^{0.4(M_{\rm v}+15)},
\end{equation} 
 where $N_{\rm gc}$ and $M_{\rm v}$ are the total number of globular clusters
in a galaxy and $V-$band absolute magnitude of the galaxy, respectively.
In the present study, we use  two different specific frequency definitions,
{\it global} and {\it local} specific frequency, both for the MPC and the MRC,
in order to compare  our present  numerical results with observations
(e.g., Kundu \& Whitmore 2001).
$S_{\rm N,P(global)}$ ($S_{\rm N,R(global)}$) and $S_{\rm N,P(local)}$ ($S_{\rm N,R(local)}$) 
represent global specific frequency of the MPC (the MRC) and local one of the MPC (the MRC), respectively.

In deriving  $S_{\rm N,P(global)}$ ($S_{\rm N,R(global)}$), 
we count total numbers of the MPC (the MRC) and estimate $M_{\rm v}$ in each model.
We adopt the following two reasonable assumptions in order to estimate 
$M_{\rm v}$: 
(1) the total stellar mass of 6.0 $\times$ $10^{10}$ $M_{\odot}$
corresponds to  $M_{\rm v}$ =  $-21$ mag
and (2) the total stellar luminosity is proportional to the total stellar mass in the model.
What should be noted here is that $M_{\rm v}$ changes during the evolution of a modelled galaxy
owing to star formation (i.e., the production of new field stars).
For example, if the total stellar mass of a merger increases by  a factor of 2.5
after merging compared with that of an initial spiral,
the final $M_{\rm v}$ is  estimated to be $-22$ mag. 
In this estimation of $M_{\rm v}$, however,
we do not  include the time  evolution of the mass-to-light-ratio  of new stars,
because the photometric properties of a merger remnant a few Gyr after 
the completion of merging is dominated by old stars, which make up 
 90 \% of the merger in mass. 
Based on $M_{\rm v}$ of a modelled galaxy,
we derive
$S_{\rm N,P(global)}$ ($S_{\rm N,R(global)}$) by using the above equation
(i.e., $N_{\rm gc}$ = $N_{\rm  MPC}$ or $N_{\rm  MRC}$, 
where  $N_{\rm  MPC}$  and $N_{\rm  MRC}$ are total numbers of the MPC and  the MRC,
respectively).
In estimating  $S_{\rm N(global)}$ (commonly accepted ``$S_{\rm N}$'')
and $S_{\rm N(local)}$ (local $S_{\rm N}$ for all globular clusters),
we assume that $N_{\rm gc}$  = $N_{\rm MPC}$ $+$ $N_{\rm MRC}$.   
We count total numbers  of stars  within a given galactocentric radius $R$ 
for  the MPC, the MRC, and all stellar components and thereby estimate the 
local specific frequency $S_{\rm N(local)}$, $S_{\rm N,P(local)}$, and  $S_{\rm N,R(local)}$.
In the initial spiral, $S_{\rm N(global)}$ and $S_{\rm N(local)}$ at $R$ = 0.5 (in our units)
are 0.8 and 0.62, respectively.

In order to discuss the formation efficiency of field  stars and globular clusters
in merging and interacting galaxies,
we introduce three quantities, $E_{\rm sf}$, 
$E_{\rm fs}$, and $E_{\rm gc}$,
which represent  the mass ratio of newly born stellar components
(i.e., field stars and globular clusters)  to initial gas,
that  of field stars  to initial gas,
and that of globular clusters to initial gas, respectively.
Below, we describe the results of 27 models and in Table 1 summarise
the model parameters for these: Model number (column 1),
total mass of a disc represented by $m_{\rm d}$ in units of $M_{\rm d}$ (column 2),
the mass ratio $m_{2}$ of two merging or interacting discs  (3),
pericentre distance $r_{\rm p}$ (4),
orbital configurations  of merging and interaction (5),
the probability of globular cluster formation $C_{\rm gc}$ (6),
$E_{\rm sf}$ (7), $E_{\rm fs}$ (8), $E_{\rm gc}$ (9),
the number ratio of the MRC to newly formed stars $N_{\rm gc}/N_{\rm sf}$ (10),
the global specific frequency $S_{\rm N}$ or $S_{\rm N(global)}$ (11),
the number ratio of MPC to MRC $N_{\rm MPC}/N_{\rm MRC}$ (12),
and comments on the models (13). 
For all models but M8 (with a gas temperature of 2500 K)
the results for a gas temperature of $10^4$ K are given.  
In the first column of the table,  ``D'', ``M'', and ``T''
represent the standard  disc model (i.e., isolated),
the merger model, and the tidal interaction one, respectively,
The model M1 is referred to as the fiducial model for convenience.
The model  with the comment of LSB (representing low surface brightness galaxies) 
has the disc's central surface brightness 
2 mag lower than that of  other models.
In addition to these 27 models, we investigate models with different $Z_{\rm g}(0)$
(i.e., initial central gaseous metallicity of the  disc)  
and with/without chemical evolution for some models.

\subsection{Limitations of the present model}

Although we investigate models with variously different parameters, 
we present the results {\it only for  representative models}
which show some typical and important behaviours of merging/interacting models
and thus help us to grasp some essential ingredients of globular cluster formation. 
The parameter dependences that are considered to be less  important 
compared with those discussed in the following sections (yet should be mentioned)
are briefly summarised as follows.
Firstly, the total number of MRC depend weakly on $P_{\rm s}$ for a plausible
range of  $2.0\times 10^4 k_{\rm B}$ $P_{\rm s}$ $2.0\times 10^6 k_{\rm B}$
in such a way that it is larger in models with smaller threshold values of $P_{\rm s}$.  
Secondly, the present results does not depend strongly on
the index of  the Schmidt law (Schmidt 1959) for a plausible
range of  1 $\le$ $n$ $\le$ 2. Thirdly initial gaseous
temperature (or pressure) is not a critically important parameter
as long as the temperature is between 2500  and  $10^4$ K.
We do not investigate the effects  of thermal feedback 
(i.e., change of internal energy of gas clouds) from supernovae
(and those of the difference in models of thermal feedback)
on the formation of field stars and clusters,
essentially because previous simulations have already
demonstrated that owing to rapid cooling of high density regions,
such thermal feedback effects do not greatly affect star formation
histories in galaxies (e.g., Katz 1992; Mihos \& Hernquist 1996).

What we should emphasise here is that we have not yet explored fully
the possible ranges of parameters of globular cluster formation.
The following parameter dependences
should be explored extensively in our future works in order that we
can confirm or modify the conclusions derived in the present study.
First is whether a possible dependence of $P_{\rm s}$ on masses of  giant
molecular clouds can change qualitatively or quantitatively the  results derived in our study. 
Second is how the formation efficiency of MRC
depends on the way to model {\it kinematic} feedback from supernovae (i.e., change of random
motion of gas clouds due to supernovae explosion).
Although this kinematic feedback effect has been demonstrated to
suppress the formation of field stars  particularly
for low-mass systems with total masses less than $10^9$ $M_{\odot}$ (Bekki \& Shioya 1999),
it is not clear how this effect is important in globular cluster formation from gas clouds 
owing to the difficulty of modelling numerically this effect on evolution
of internal structures  of molecular clouds. 
Third is how the detailed  processes of globular cluster formation in gas clouds
under high gas pressure depend on
the differences in physical properties of individual (molecular) gas  clouds
(e.g., internal structures, temperature, and the strength of magnetic field).
The present galaxy-scale simulations, 
which can not resolve the scale down to sub-pc 
(where local hydrodynamical processes determine the clouds' properties) 
do not enable us to discuss this important point. 
The present model is not so sophisticated as to include all of these three
important effects in numerical simulations. Therefore it should be noted 
that our numerical simulations should be carefully interpreted owing to these yet
unexplored possible important effects (which can possibly modify  the present results)
and can be served 
as a first approximation model of actual globular cluster formation.
 
\begin{figure}
\psfig{file=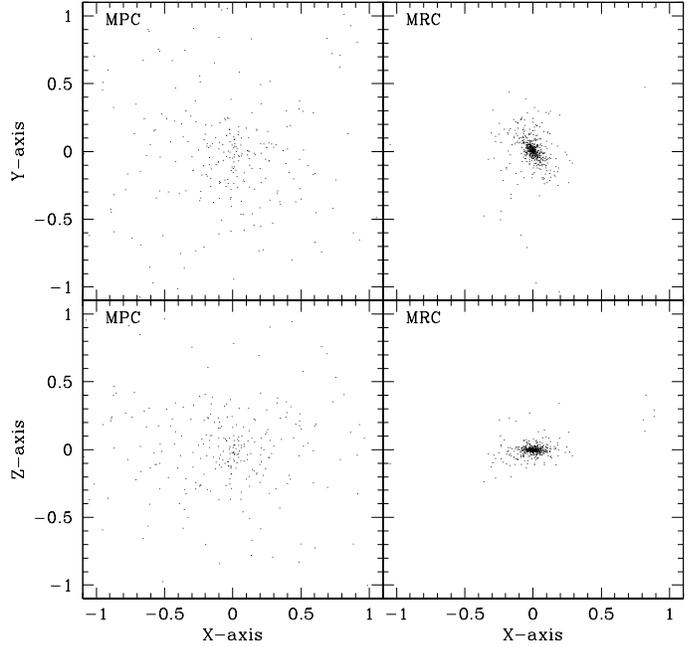,width=9.cm}
\caption{ 
Final mass distribution of the MPC (left two panels) and the MRC (right)
projected onto the $x$-$y$ plane (upper two)
and onto the $x$-$z$ plane (lower) in the fiducial model M1 at $T$ = 14 ($\sim$ 2 Gyr).
Note that the final  distribution is more centrally concentrated for   the MRC
than  the  MPC.
}
\label{Figure. 3}
\end{figure}

\begin{figure}
\psfig{file=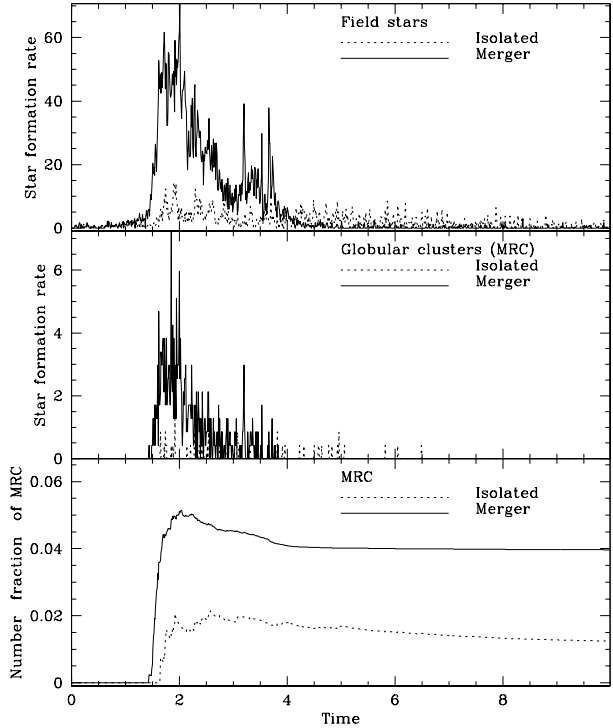,width=8.cm}
\caption{ 
Time evolution of star formation rate (in units of $M_{\odot}$ ${\rm yr}^{-1}$) 
of  field stars (top) and metal-rich globular clusters represented by the MRC (middle)
both for the standard isolated disc model D1 (dotted lines) and for the fiducial merger
model M1 (solid). The bottom panel shows the time evolution
of the  ratio of the total number of the MRC to that of newly born stars 
(i.e., new field stars and the MRC) for the two models, D1 and M1.
Note that the peak of the MRC formation rate is nearly coincident with that of the field star
formation rate in the merger model and that the duration of  MRC formation
is shorter than that of field stars: the MRC formation starts later than and ends
earlier than the field star formation.
The number of MRC formed in a merger greatly exceeds an isolated galaxy
}
\label{Figure. 4}
\end{figure}

\begin{figure}
\psfig{file=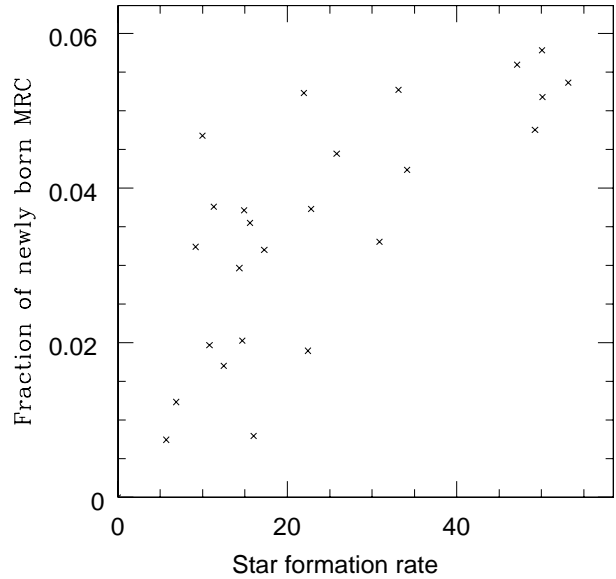,width=8.cm}
\caption{ 
A correlation between 
number fraction of newly formed the MRC among all newly born stellar components 
(i.e., new field stars and the MRC) 
and 
star formation rate (in units of $M_{\odot}$ ${\rm yr}^{-1}$)
in the merger model M1. 
Each data point represents the mean star formation rate averaged over
0.1 $t_{\rm dyn}$ (corresponding to 14.1 Myr in this model) at that  time step
and the number (or mass) ratio of the MRC (formed within 0.1 $t_{\rm dyn}$ at that time  step)
to newly born  stellar components.
Therefore the positive correlation 
implies that gas 
is more likely to be  converted into MRC rather than into   field stars
for a galaxy merger with a high star formation rate.
}
\label{Figure. 5}
\end{figure}

\begin{figure}
\psfig{file=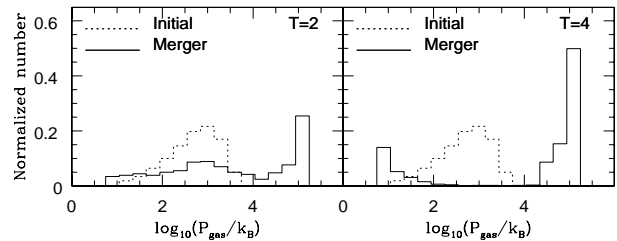,width=8.cm}
\caption{ 
Distribution (by number) of gas pressure (in units of $k_{\rm B}$) at $T$ = 2 (left)
and $T$ = 4 (right) in the merger model M1. For comparison, the result of the initial discs
is also given.
In this histogram,
the gaseous pressure $P_{\rm gas}$ is estimated for each SPH gaseous particle.  
Note that 
the merger includes a larger amount of  high pressure gas with
$P_{\rm gas}$ $\sim$ $10^5$ $k_{\rm B}$.
This is essentially because
the strong tidal forces of galaxy merging form  high-density shocked 
gaseous regions in the merger. 
}
\label{Figure. 6}
\end{figure}

\begin{figure}
\psfig{file=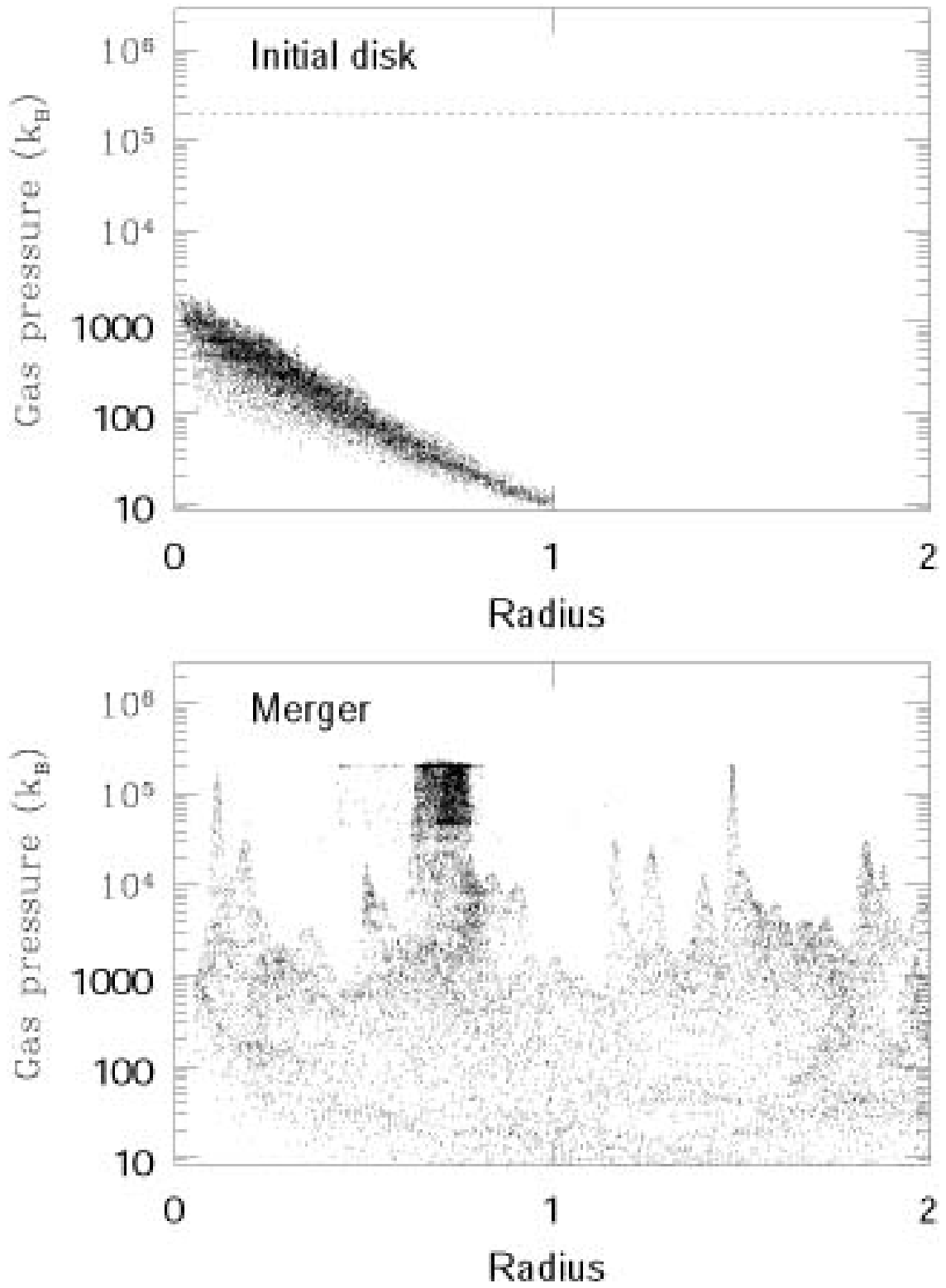,width=8.cm}
\caption{ 
$Upper$:  Distribution of gaseous particles  
on a pressure-radius plane  for the initial  disc model D1.  
The dotted line  shows the threshold pressure over which
globular clusters  are assumed to be formed. 
$Lower$:  Distribution of  particles that are $initially$ $gaseous$ $particles$  
on a pressure-radius  plane  
for the major merger model M1 
at $T$ = 2 (0.28 Gyr). 
Here the ``radius'' means the distance (in our units, i.e., 17.5 kpc in this model)
from the centre of mass of the merger. 
Not only gaseous particles but also new stellar ones  (the MRC and field stars)
formed before $T$ = 2 are plotted.
For each of these new stellar components,
gas pressure  at the epoch when the precursor
gaseous particle is converted into the new stellar one is plotted. 
Accordingly, by comparing the upper panel 
with the lower one, we can clearly observe
how drastically global dynamical evolution
of major galaxy merging has increased the gas pressure. 
The regions around the radius of 0.7 in our units ($\sim$ 12 kpc)
 represent the central star-burst cores
in the merger. 
}
\label{Figure. 7}
\end{figure}

\section{Results}
\subsection{Merging}
\subsubsection{Formation and  evolution of globular  clusters}

Figs. 1 and 2 summarise the dynamical evolution of old stars, new field stars, the MPC, and the MRC
in the fiducial major merger model M1.
As dynamical relaxation of major merging proceeds, strong non-axisymmetric forces
due largely to stellar bars efficiently drives  a large amount of gas into
the central regions of the two discs, consequently triggering  a strong  star-burst
in the central regions of the merger.
New field stars (i.e., star-burst populations)  form a  young, metal-rich
core with an  effective radius of 0.05 ($\sim$ 0.88 kpc, a factor of $\sim$ 5 smaller
than that of the old stellar component)
in the remnant elliptical galaxy.
The new field stars  show arc-like structures  
even after the completion of merging ($T$ = 14 corresponding to 2 Gyr).
Several tidal dwarf galaxies composed of old stars, gas, and new
field  stars are formed during merging, the most massive of which
survives  tidal destruction and orbits the final merger remnant elliptical.
Orbital angular momentum of the merging  discs is rapidly
converted into internal spin of the single merger remnant owing
to frictional drag of the tidal force of the merger. 
As a natural result of this angular momentum redistribution,
old stars initially located in the  outer regions of the discs
finally form the outer stellar halo with a relatively large amount
of intrinsic angular momentum in the formed  elliptical galaxy. 
These results are basically consistent with the numerical simulations
of dissipative major merging of Mihos \& Hernquist (1996).

The two  MPC systems, which are located in the halo regions of the spirals,
suffer more severely from violent relaxation of major merging
and consequently form a single  MPC system surrounding the elliptical galaxy. 
Owing to the  angular momentum redistribution,
the effective radius of the MPC  becomes larger (0.61 in our units, corresponding to 10.7 kpc) 
compared with the initial disc model (5.2 kpc).
The radius within which 90 \% of the MPC (old stars) are  included is also dramatically changed
from 1.27 (0.61) into 2.32 (1.52) in our units during merging.
This implies that a significant  fraction of the MPC are  tidally stripped from the merger,
and also that the MPC have a  more extended distribution than background old stellar components. 
These results imply that as a galaxy experiences  multiple   major  mergers,
the half-mass radius of  the MPC 
becomes progressively larger.
The MRC, on the other hand, can remain in the central region of the merger,
essentially because they are initially formed in the merger's central region,
where outward angular momentum transfer (or tidal stripping) is less strong.
Unlike the MPC (which follow only dissipationless dynamics), 
the MRC follow dissipative dynamics of the gas before their formation
and dissipationless dynamics  afterwards.  
As a result of this, both the effective radius of the MRC
and the radius within which 90 \% of the MRC are included 
are small,  0.05 (0.88 kpc) and 0.2 (3.5 kpc), respectively,   at $T$ = 14. 
We find only a  small fraction of the MRC are transferred to the outer region
of the merger remnant.
As is shown in Fig. 3,
the MRC distribution is much more flattened and centrally concentrated than
that of the MPC. 
Thus, these results suggest that the difference in final spatial
 distributions between the MPC and the MRC 
can be caused by the difference in dynamical evolution between the MPC and the MRC in major mergers.

Fig. 4 shows that galaxy merging dramatically increases 
not only the star formation rate of field stars,  but also
that of the MRC (compare the results of the isolated model and those of the merger one). 
What should be emphasised here is that the burst of the MRC formation starts later than,
and ends earlier than that of the field stars in the merger model, i.e.,
the duration of the MRC formation is shorter than that of the field stars.
Furthermore, the ratio  of the MRC to  new stellar components
(i.e., new field stars plus  the MRC) rapidly becomes large ($\sim$ 4 \%) and
this is a more dramatic effect in the merger model than in the isolated one. 
These results indicate that gas is more likely to be converted into
the MRC rather than field stars during major merging, since the
fractional amount of stellar light coming from the MRC among all
young stars and  clusters is higher in major mergers with star-bursts
than in quiescent  galaxies.
As demonstrated in Fig. 5, the available gas is more likely  
to be converted into the MRC rather than new field stars 
when the star formation rate is higher.
Indeed, we confirm that this strong positive correlation between
the star formation rate and the formation efficiency of the MRC
is true for the majority  of our models.
Recently, van den Bergh (2001) discusses  that the fraction of $U$-band light
of galaxies that is generated by young clusters is proportional to the rate of
star formation per unit area, based on the observational data by
Larsen \& Richtler (2001).
This observational result is at least qualitatively consistent with
the above numerical results.

The origin of the increased efficiency of the formation of the MRC
during the star-burst is highlighted  in figs. 6 and 7.
Fig. 6 demonstrates that the number fraction of gas with $P_{\rm gas}$ $>$ $10^5$ $k_{\rm B}$
in the merger model is greatly increased due to  the strongly shocked gaseous regions,
reaching  25 \%  at the strong star-burst epoch of $T$ = 2 and 50 \% at $T$ = 4.
Furthermore, the overall distribution of gas pressure becomes bimodal at $T$ = 4
with the relatively  low gas pressure being due to the tidally stripped halo gas. 
Fig. 7 also confirms that the pressure of some gas particles,  particularly in the central regions
of the merger, becomes more than two orders of magnitude higher than that in the initial
disc. 
Thus,
the MRC are formed  more efficiently from gas in major merging owing to this larger amount of
gas with high pressure exceeding the threshold value for the collapse of molecular clouds. 
These results clearly explains why young star clusters or super star clusters
are observed to be located preferentially in interacting and merging galaxies
with strong star-bursts.
The high gas pressures during major mergers can collapse 
not only low-mass clouds, but also those which are significantly
more massive. This leads to a bias in the mass function
of proto-globular clusters towards higher masses than is the
case for star clusters formed in lower-pressure environments.

The physical reasons for the derived high gas pressure are the following two.
First is that owing to 
rapid and efficient radial gas inflow induced by non-axisymmetric
structures formed in merging  and by gas dissipation,
the merger forms the central very high density gas regions.
The high density regions correspond to high pressure ones
because of the adopted equation of state for gas.
Second is that non-axisymmetric perturbations of merging
forms strong tidal tails,  where gas can be  dramatically compressed
(and energy  dissipation proceeds very efficiently) 
so that it becomes very high density and thus high pressure. 
These two processes do not depend on model parameters adopted
in this study, although it depends on merger orbital parameters 
whether MRCs are formed more efficiently in the compressed tidal tails 
or in the central regions of mergers.
Thus high gas pressure induced by dissipative dynamics of galaxy merging
can be considered  to be essential for MRC  formation.
The apparent upper limit of gas pressure in Fig. 7 is that
if $P_{\rm gas}$ $>$ $P_{\rm s}$,
gas particles are immediately  converted into stars
and consequently can not contribute to the increase of gas pressure. 

\begin{figure}
\psfig{file=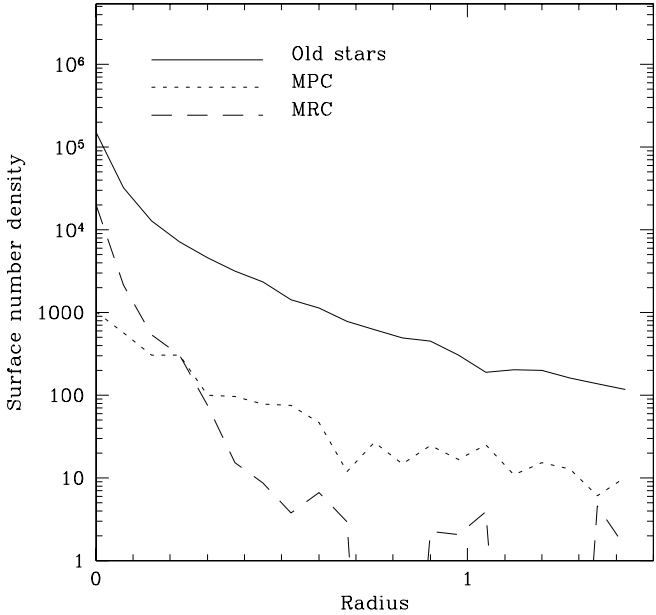}
\caption{ 
The radial surface number density distribution both for the MPC (dotted) and the MRC (dashed)
in the merger model M1 at $T$ = 14 ($\sim$ 2 Gyr).
The radius is given in our units (17.5 kpc) and the result for old  stars
initially within the two spirals are also plotted by a solid line for comparison.
In this model, 1000 in the ordinate axis corresponds to $\sim$ 3.27 number ${\rm kpc}^{-2}$.
}
\label{Figure. 8}
\end{figure}


Fig. 8 shows that the surface number density distribution is very different 
between the two globular cluster populations for the elliptical
galaxy formed by major merging. 
The MRC show  steeper and more centrally concentrated profiles than the MPC.
This is primarily because the MRC are formed from gas that experiences gas 
dissipation before they are converted into stellar components
whereas the MPC experiences only dissipationless dynamical relaxation 
that can not significantly increase the central density of the MPC.
The  MPC distribution  for $R$ $>$ 0.5 is appreciably flatter 
than that of the old stellar components,
which is consistent with observations.
Fig. 9  shows that the number ratio of the MRC is higher in the inner regions
of the elliptical galaxy, which reflects the fact that the MRC are formed from
gas that is rapidly  transferred into the central region of the merger
because of efficient gas  dissipation.
These  differing spatial distributions 
 imply that the overabundance of the MRC in the central regions
of elliptical galaxies 
can result in the formation of negative metallicity (thus colour) gradients 
of globular cluster systems.
The  difference in the surface number density distribution
between the MPC and the MRC is consistent at least qualitatively
with observational results.

\begin{figure}
\psfig{file=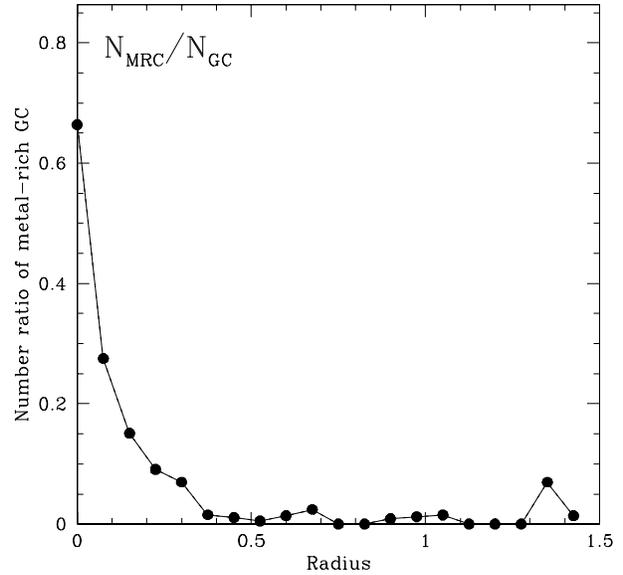,width=8.cm}
\caption{ 
Radial dependence of the ratio of total number of the MRC to that  of the MPC and the MRC
in the merger model M1. 
The radius is given in our units (17.5 kpc).
}
\label{Figure. 9}
\end{figure}

As a natural result of this difference in the spatial distribution,
the radial dependence of local specific frequency  
becomes qualitatively different between the MPC and the MRC.
Fig. 10 shows  that $S_{\rm N, P(local)}$ increases  weakly with radius 
whereas $S_{\rm N, R(local)}$ decreases  with radius.
This is essentially because the MPC are more spatially extended 
than the (old) field stars  due to 
the outward transfer of the MPC during merging whereas
the MRC are more centrally concentrated than the field stars
due to  their dissipative formation. 
Since the final total number of the MRC is larger than that of the MPC,
this results in a  decrease  of $S_{\rm N(local)}$ 
with radius.
Although the derived $S_{\rm N(global)}$ is well within 
the observed value of field elliptical galaxies,
the radial dependences of $S_{\rm N, R(local)}$ and $S_{\rm N(local)}$
are apparently inconsistent with observations.
This inconsistency means either that
the merger scenario of globular cluster formation
has a serious problem that has not been pointed out by Ashman \& Zepf (1992)
or that  the MRC are not so efficiently formed in the central region
as the present model predicts.
We here point out that this inconsistency can be mitigated if
we consider the tidal destruction of MRC in the central region of mergers.
Although investigating the tidal destruction of MRCs in galaxy mergers   
is important for better understanding reasons for the inconsistency between
our simulations and observations, 
this is our future study because of the lack of numerical resolution
in the present study.
We will discuss this problem later in \S 4.

\begin{figure}
\psfig{file=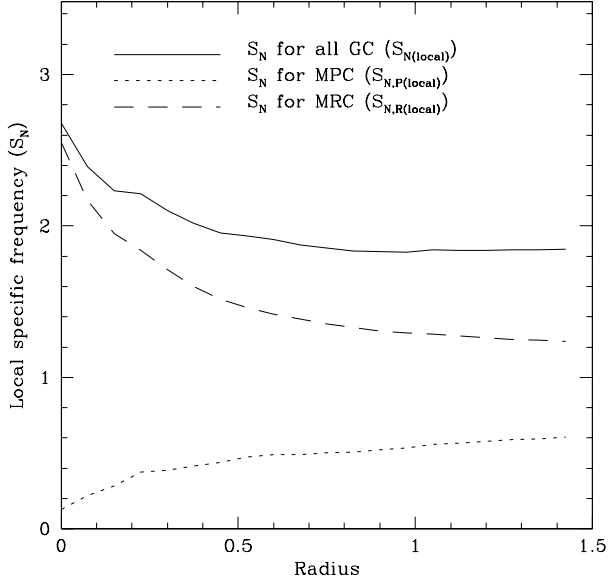,width=8.cm}
\caption{
The radial dependence of specific frequency ($S_{\rm N}$) for all globular clusters
(represented by solid line; $S_{\rm N(local)}$), the MPC (dotted;$S_{\rm N,P(local)}$),
and the MRC (dashed; $S_{\rm N,R(local)}$).
The radius is given in our units (17.5 kpc).
}
\label{Figure. 10}
\end{figure}

\begin{figure}
\psfig{file=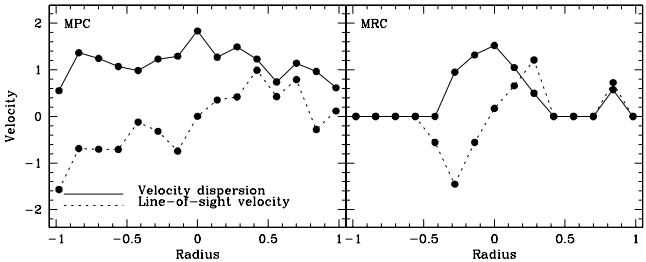,width=8.5cm}
\caption{
Kinematical properties of the MPC (left) and the MRC (right).
The line-of-sight velocity (seen from $y$-axis) at each radius
and the velocity dispersion are represented by 
dotted lines and solid ones, respectively.
The radius and velocity are given in our units (17.5 kpc and 121 km s$^{-1}$, respectively).
}
\label{Figure. 11}
\end{figure}

As is shown in Fig. 11, 
the MPC has a larger velocity dispersion in the central regions 
and a larger line-of-sight velocity in the outer galaxy regions.
Since the MPC  are assumed to have
no  global rotation with respect to the  initial spirals, 
the derived larger  line-of-sight velocity  of the MPC
suggests that 
the MPC spin up during merging,  owing to the conversion of orbital
angular momentum of the merging spirals into intrinsic angular momentum
of the merger remnant
(because they are initially located in the spirals outer part
where angular momentum conversion or transfer is very efficient).
This result is consistent with the prediction by
Ashman \& Zepf (1992) and Zepf \& Ashman (1993).
The MRC shows the same kinematical properties   as  those of the MPC for $R$ $<$ 0.3.
In this prograde-prograde merger model, both the MPC and the MRC show global
rotation.

\begin{figure}
\psfig{file=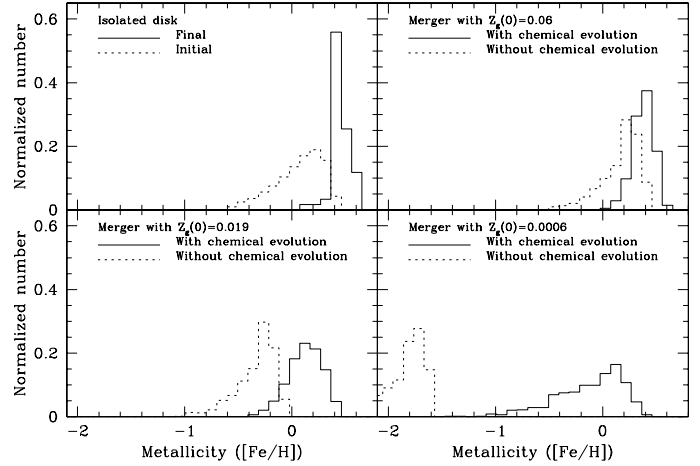,width=9.cm}
\caption{
Metallicity distributions for the isolated disc model D1 (upper left) and the merger models
M1 with  different values of the initial central gaseous metallicity represented
by $Z_{\rm g}(0)$.
For $Z_{\rm g}(0)$ = 0.06 (upper right),
the gaseous metallicity at solar radius (at $R$ = 0.49 in our units
corresponding to  8.5 kpc) is 0.02. Accordingly this model
is reasonable for the present-day disc-disc mergers.
The  models with $Z_{\rm g}(0)$ = 0.019 (lower left)
and 0.0006 (lower right) can be considered to be higher redshift mergers
because of their low metallicities. 
For comparison, 
results  without chemical evolution
are also given in dotted lines. 
}
\label{Figure. 12}
\end{figure}

Fig. 12 shows final metallicity distributions of the MRC in the fiducial merger model M1
with different initial central gaseous metallicities ($Z_{\rm g}$(0) = 0.06, 0.0019, and 0.0006)
and the standard isolated spiral  model D1.
Fig. 12  shows  the metal-rich peak of the MRC in the merger remnant.
It is  clear from this figure that the derived peak metallicity of the MRC
in the model with $Z_{\rm g}$(0) (corresponding to a merger at the redshift $z$ = 0)
is [Fe/H] $\sim$ $+0.4$,  much higher than the observed mean value
of [Fe/H] $\sim$ $-0.5$ in elliptical galaxies.
This result does not depend  strongly on the initial gaseous metallicity gradients
of two spirals.
The physical reasons for this rather high metallicity are    that
the progenitor discs are assumed to have initial high  metallicities  ([Fe/H] $\sim$ $+0.17$)
and  that the MRC are formed from gas preferentially in the merger's central region,
where gas metallicity is  high owing to the adopted negative metallicity gradient
and, 
chemical enrichment proceeds rapidly  due to  the central star-burst. 
If we assume that  the MRC are formed only from the initial relatively metal-poor
gas for some reason (e.g., delayed chemical evolution or no  MRC formation after
the first generation of massive stars in the MRC
due to destruction processes of giant molecular clouds, such as strong thermal
and kinematical feedback effects),
the peak is still higher than  the observed value in elliptical galaxies
(see the results for the model with $Z_{\rm g}$(0) = 0.06 without chemical evolution in Fig. 12).
The model with moderately low initial gaseous metallicity and without chemical evolution
or that with very low initial gaseous metallicity and with chemical evolution
can provide the peak metallicity [Fe/H] of the MRC  between $-0.5$ and 0. 
These results imply that the present-day gas-rich disc mergers
can not explain the observed metallicity distributions of the MRC in ellipticals.
Therefore if most  elliptical galaxies are formed by major merging,
then it  must occurred at early epochs.
This conclusion does not depend on merger orbital parameters.
The derived  MRC with {\it supersolar metallicity} in these models are inconsistent
with observations, which implies that such MRCs (most of which are located in
the central regions of mergers) 
can be preferentially destroyed
in the central cores of mergers because of strong tidal force.
Fig. 13 demonstrates that the MRC have a negative metallicity gradient
with a  slope similar to that of new  field stars.
The reason for this is firstly that the initial metallicity
gradient of gas forming the MRC can not be completely smoothed out 
even after violent relaxation of major merging and secondly that
the gas dissipation and the chemical enrichment
cause  the gradient to be steeper.

\begin{figure}
\psfig{file=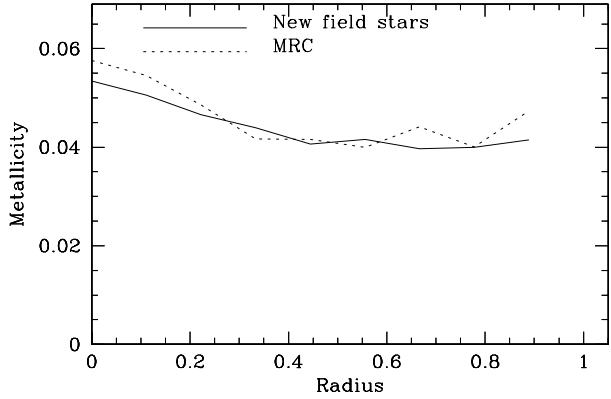,width=8.cm}
\caption{ 
The radial metallicity gradient of new field stars (solid) and the MRC (dotted)
in the merger model M1.
The radius is given in our units (17.5 kpc).
}
\label{Figure. 13}
\end{figure}

The present major merger models furthermore demonstrate that
a few dwarf galaxies with a few globular clusters 
can be formed
during merging (see Figs. 1 and 2) and also that  only one
of them (seen at $(x,y)$ $\sim$ $(2.0,1.0)$ in figs. 1 and 2) 
can survive from the violent relaxation.
This result confirms the earlier numerical results
by Barnes \& Hernquist (1992) and provides the following implication
on globular clusters in dwarf galaxies.
The so-called `tidal dwarf' formed during merging can contain
very bright young globular clusters. 
These young globular clusters in a tidal dwarf might well  evolve into a single
giant globular cluster or a galactic nucleus owing to efficient
dynamical friction in the dwarf and the resultant merging of the clusters.
Formation of a galactic nucleus (or giant globular clusters like $\omega$Cen)
in a dwarf, which has already been demonstrated by Oh, Lin, \& Richer (2000),
can be responsible for the observed very massive  clusters in some merger remnants 
(e.g., Goudfrooij et al. 2002). 
We suggest that some of giant globular clusters in merger remnants can be embedded by
the very low surface-brightness envelopes of their host dwarf galaxies.
Future very deep imaging of the surrounding regions of very massive globular
clusters in merger remnants are important to see whether the giant clusters (like W3 in NGC 7252)
are embedded by dwarf envelopes.

\begin{figure}
\psfig{file=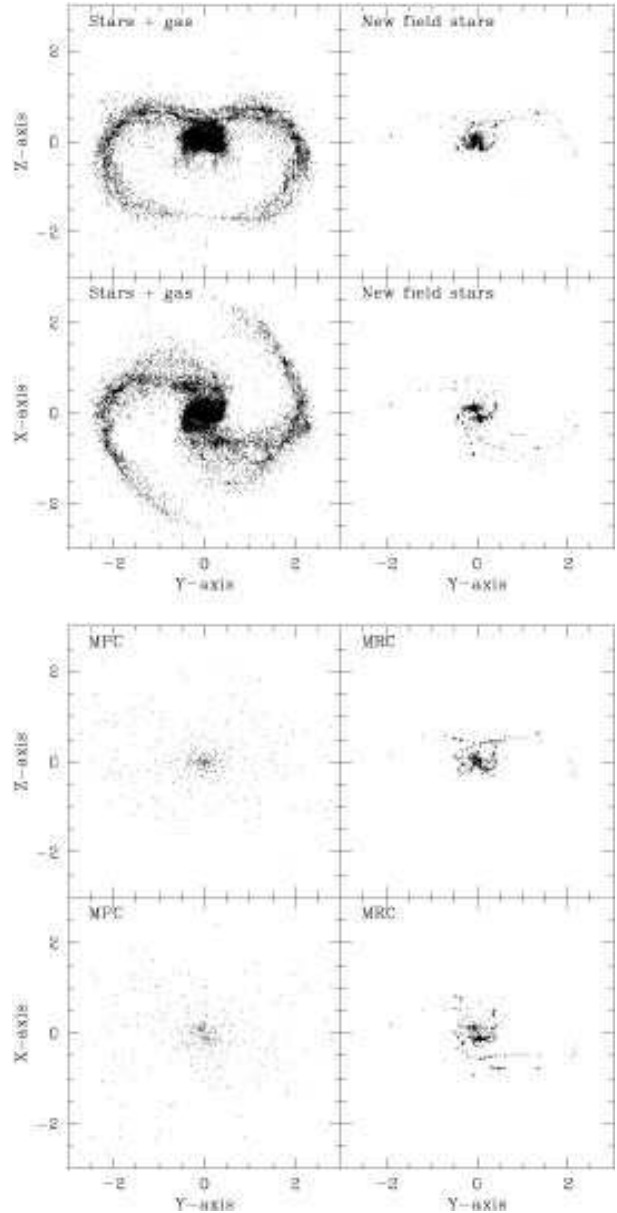,width=8.cm}
\caption{ 
{\it Upper four:} Mass distribution of old stars and gas  (left two panels) 
and new field stars  (right)
projected onto the $y$-$z$ plane (upper two)
and onto the  $y$-$x$ one (lower) 
in the Antennae  model M8 with $C_{\rm gc}$ = 1.0
at $T$ = 4 ($\sim$ 0.56 Gyr).
{\it Lower four:} 
The same as the upper four but for
the MPC (left two panels) and the MRC (right).
}
\label{Figure. 14}
\end{figure}

\begin{figure*}
\psfig{file=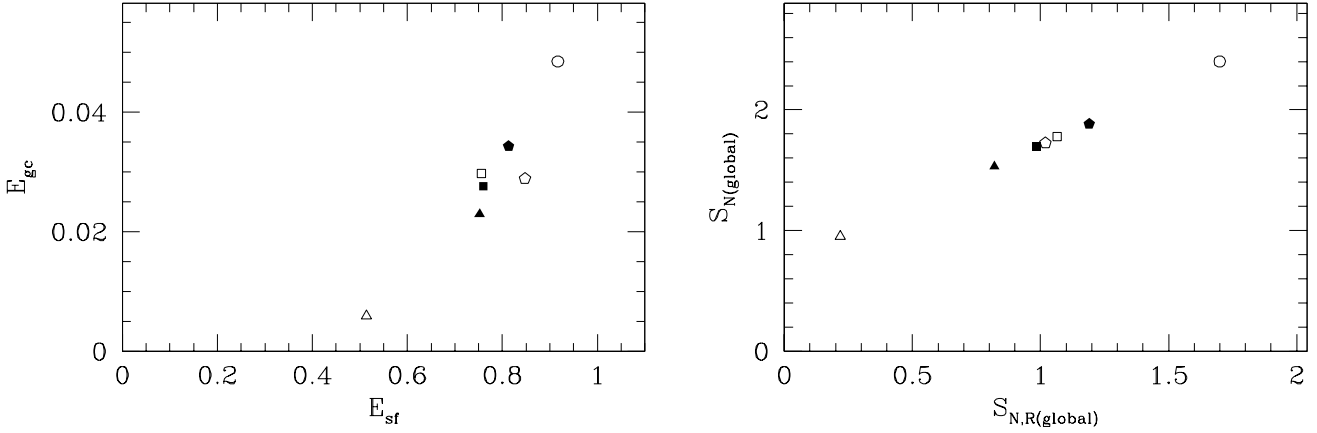}
\caption{
Distribution of the six merger models (M1, M2, M3, M4, M5, and M6)
with different orbital configurations and the isolated disc model D1 on
$E_{\rm sf}$-$E_{\rm gc}$ plane (left) 
and on $S_{\rm N,R(global)}$-$S_{\rm N(global)}$ plane (right).
$E_{\rm sf}$ and $E_{\rm gc}$ describe the formation efficiency
of newly born stars (total number of new stars divided by that of initial gas)
and that of metal-rich globular clusters  (total number of the MRC divided by
that of initial gas), respectively.
D1, M1 (fiducial), M2 (prograde-retrograde), M3 (retrograde-retrograde), 
M4 (highly inclined discs), M5 (larger orbital angular momentum), and  M6 
(higher gas mass fraction) are represented by
open triangle, open square, open pentagon, open circle,
filled triangle, filled square, and filled pentagon, respectively. 
Note than globular cluster formation
is more efficient in a merger showing higher efficiency of star formation.
Note also that a merger with higher $S_{\rm N,R(global)}$
shows higher $S_{\rm N(global)}$.  
}
\label{Figure. 15}
\end{figure*}

\begin{figure*}
\psfig{file=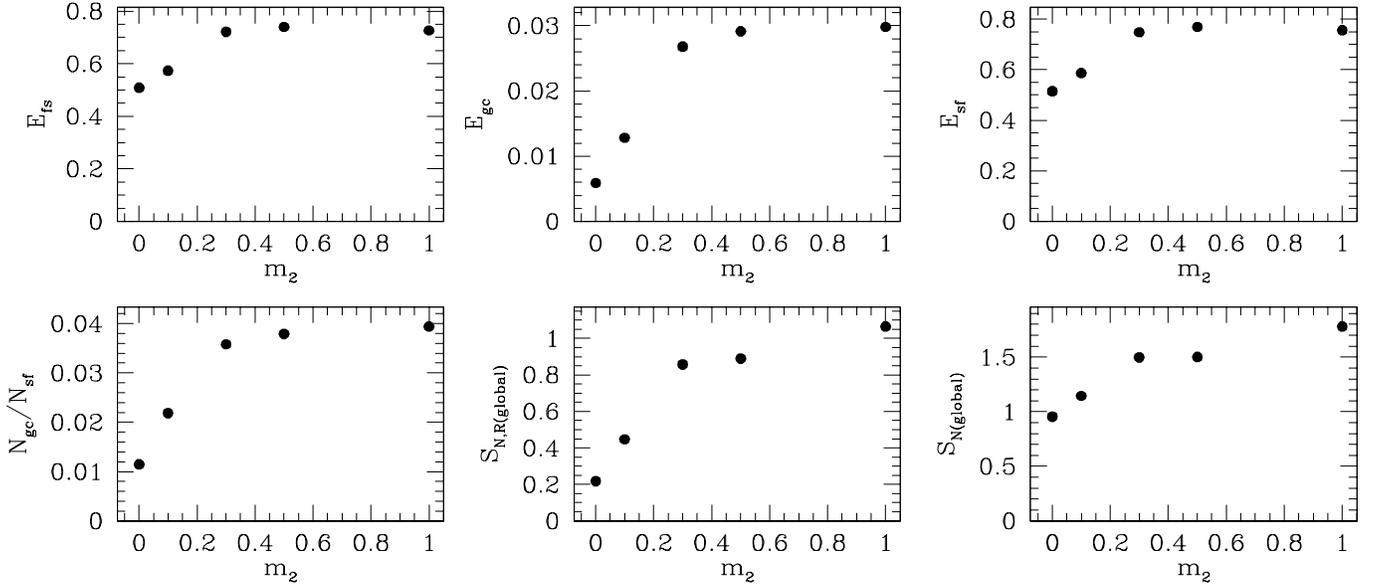,width=18.cm}
\caption{ 
The dependence of  $E_{\rm fs}$ (upper left),
$E_{\rm gc}$ (upper middle), $E_{\rm sf}$ (upper right),
$N_{\rm gc}/N_{\rm sf}$ (lower left),   $S_{\rm N,R(global)}$ (lower middle),
and $S_{\rm N(global)}$ (lower right) on the ratio of two merging discs
(represented by $m_{2}$) for the merger models M1, M11, M12, and M13.
For comparison, the isolated disc model D1 ($m_{2}$ = 0) is also plotted.
Physical meaning of the above six quantities are given in the text. 
}
\label{Figure. 16}
\end{figure*}

\begin{figure}
\psfig{file=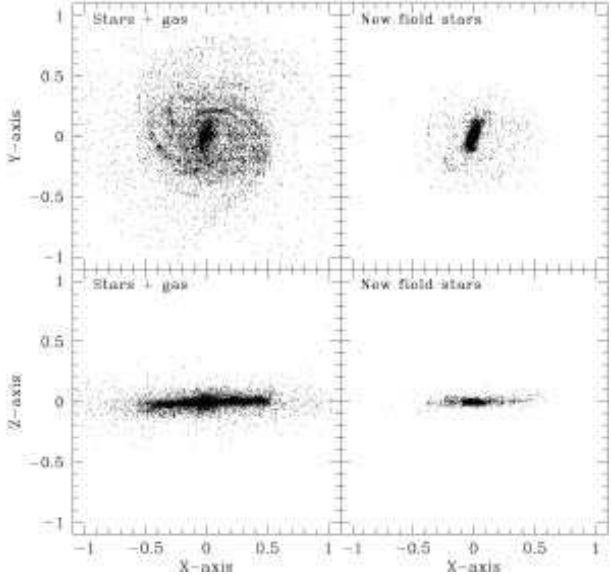,width=8.cm}
\caption{ 
Mass distribution of old stars and gas  (left two panels) 
and new field stars  (right)
projected onto the $x$-$y$ plane (upper two)
and onto the  $x$-$z$ one (lower) 
in the tidal interaction  model T1
at $T$ = 8 ($\sim$ 1.13 Gyr).
}
\label{Figure. 17}
\end{figure}

\begin{figure}
\psfig{file=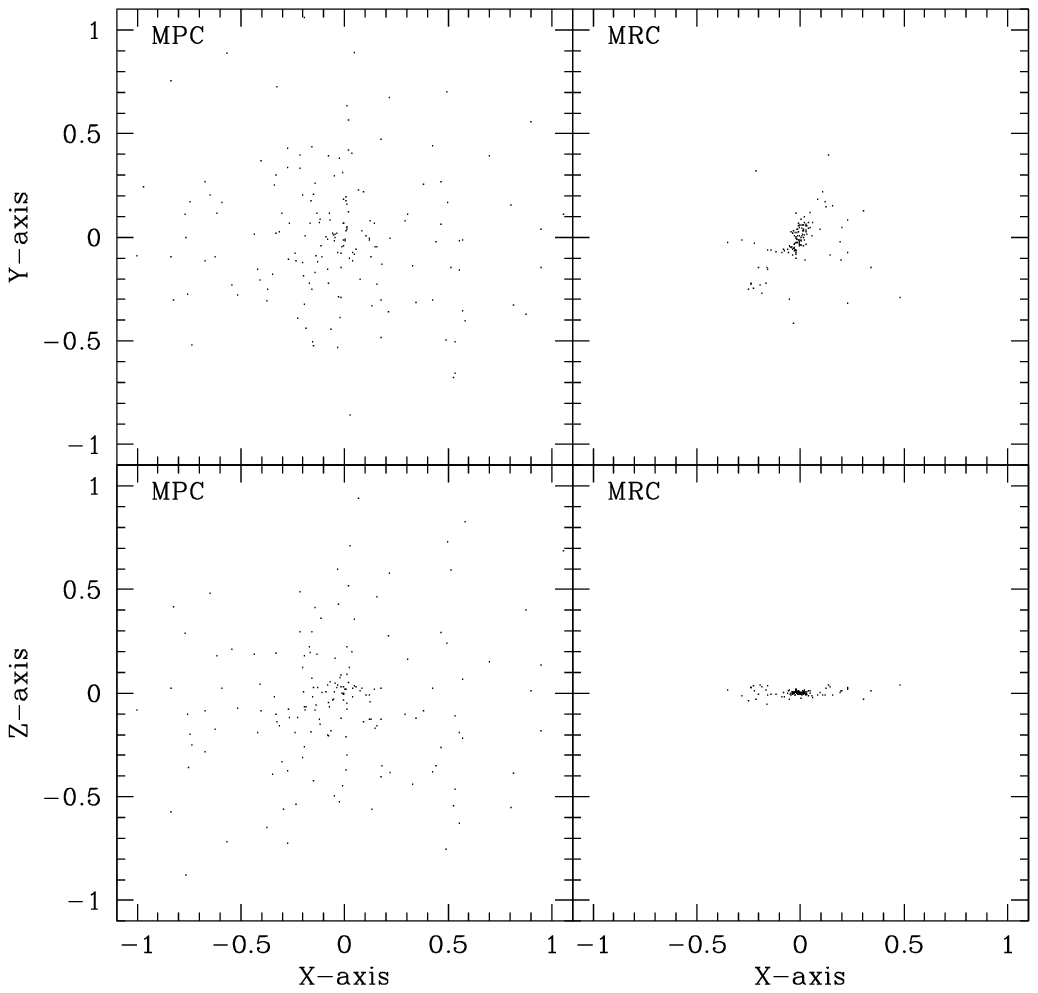,width=8.cm}
\caption{ 
The same as Fig. 17 but for the MPC (left) and the MRC (right).
}
\label{Figure. 18}
\end{figure}

\subsubsection{Dependence on initial conditions of merging}
Although the numerical results 
of structure, kinematics, and chemical properties
of the MPC and the MRC are similar to  one another 
between the fiducial model M1 and  other  merger models,
these depend on (1) initial conditions of mergers such as orbital configurations
of merging and (2) the mass-ratio of merging two spirals ($m_{2}$).
In Figures  14,  15,  and 16,
we illustrate  the derived dependences on initial conditions and mass ratio. 
 We find the following:

 (i) The range in orbital configurations of merging introduces
the diversity in $E_{\rm gc}$, $E_{\rm sf}$,
$S_{\rm N(global)}$, and $S_{\rm N, R(global)}$.
In particular, 
the observed range  of $S_{\rm N(global)}$ in elliptical galaxies
can be due  partly to the difference in orbital configurations of major galaxy mergers
that forms elliptical galaxies.
However, there are correlations between these, i.e., 
$E_{\rm gc}$ is proportional to $E_{\rm sf}$ whereas $S_{\rm N(global)}$
is proportional to $S_{\rm N,R(global)}$.

   (ii) The retrograde-retrograde merger model (M3) 
shows both higher  efficiency of the MRC (higher $E_{\rm gc}$) and $S_{\rm N,R(global)}$ 
(thus  $S_{\rm N(global)}$)
compared with other models (e.g., M2, M4, and M5).
This implies that elliptical galaxies with kinematically  distinct cores,
in particular, with counter-rotating cores 
(which are demonstrated  by Hernquist \& Barnes 1991
to be  formed by retrograde-retrograde mergers)
are likely to have higher $S_{\rm N,R(global)}$.\\
\indent (iii) The model with higher 
initial gas mass fraction of $f_{\rm g}$ = 0.2 (M6) shows appreciably
higher $S_{\rm N(global)}$ compared with the fiducial model M1 with $f_{\rm g}$ = 0.1, 
which implies that elliptical galaxies formed at  higher redshift will  show higher 
$S_{\rm N(global)}$.
 
(iv) Both  $E_{\rm gc}$ and $S_{\rm N(global)}$ ($S_{\rm N,R(global)}$)
depend on  $C_{\rm gc}$ in such a way that a merger with a larger
$C_{\rm gc}$ shows higher $E_{\rm gc}$ and $S_{\rm N(global)}$
(compare these values of M1 with those of M9 and M10 in  Table 1).
Furthermore, in the model with the higher  $C_{\rm gc}$,
the MRC can be efficiently formed not only in the central region,  but also
in the shocked gaseous regions along tidal tails.
Fig. 14 describes  an example of this (the results of the Antennae model M8), 
showing the MRC formation along the two  tidal tails.
This the MRC formation along the tails
can not be clearly seen in the Antennae model M7 with  $C_{\rm gc}$ = 0.1.
The morphology of the MRC in  the  model M8 is similar to the observed one
in NGC 4038/39, which implies that the formation efficiency
of the MRC along tidal tails is rather high.
These results also provide an explanation for the origin
of young clusters along tidal tails observed in a compact group of galaxies
(``the Stephan's Quintet'') which includes interacting/merging galaxies
(Gallagher et al. 2001).
Olson \& Kwan (1990) demonstrated that enhanced cloud-cloud
collision rates during merging can lead to the cloud coalescence,
which is responsible for the formation of very massive clouds.
High gas pressure derived in the present merger models
might well collapse these super massive gas (molecular) clouds. 
Therefore  the present study suggests that
the origin of the observed very bright young clusters such as the knot ``S''
in the tidal tail of the Antennae (Whitmore et al. 1999) 
is  closely associated with the collapse of super-giant molecular clouds 
(by high gas pressure)
formed in  galaxy merging with high $C_{\rm gc}$ due to coalescence of giant  molecular clouds.

(v)  $E_{\rm fs}$,  $E_{\rm gc}$, $E_{\rm sf}$,  $N_{\rm gc}/N_{\rm sf}$,
$S_{\rm N,R(global)}$,  and $S_{\rm N(global)}$ are all likely to be larger
for the  model with the  larger $m_{2}$.  
This result means that the stronger tidal force from  galaxy merging in  the model with
the larger $m_{2}$ can convert a larger amount of gas more efficiently
into new field stars and  the MRC.
Considering that the final morphology of a merger depends strongly on $m_{2}$
of the merger's two spirals such that the major merger with $m_{2}$ $\sim$ 1 
becomes an elliptical galaxy whereas 
the unequal-mass (or minor) merger with $m_{2}$ $\sim$ 0.3 (0.1)
becomes an S0 (e.g., Bekki 1998),
the above dependence suggests a correlation between $S_{\rm N(global)}$ and
Hubble morphological types of galaxies.
To be more specific, elliptical galaxies can  have higher $S_{\rm N(global)}$ than
S0s and S0s can have higher $S_{\rm N(global)}$ than late-type disc galaxies.

(vi) Irrespective of merger parameters,
most MRC can be seen in the central regions of merger remnants.
This is consistent with the observed young clusters
centred on merger remnants such as NGC 7252 (Miller et al. 1997),
NGC 3256 (Zepf et al. 1999), and NGC 3921 (Schweizer et al. 1996).
However such central concentration of the MRC in a merger remnant causes the larger $S_{\rm N}$
in the inner region of the  merger, which is inconsistent with observation. 
These MRC in the central region of a merger remnant
can be responsible for the formation of a compact nucleus
or a compact cluster of the MRC 
through dynamical friction of the MRC (Oh et al. 2000; Bekki \& Couch 2001).
   
\begin{figure*}
\vspace{1.cm}
\psfig{file=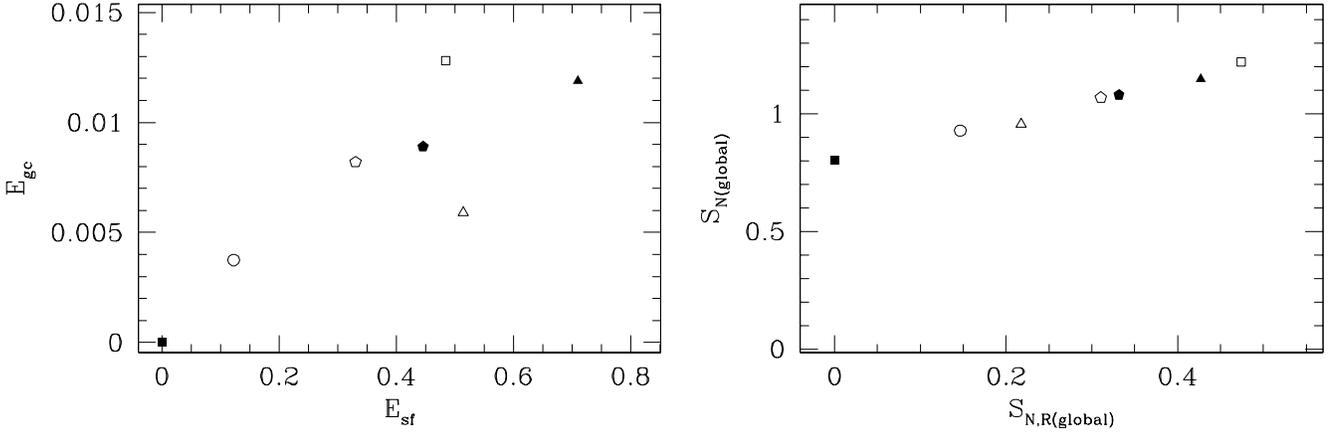}
\caption{ 
Distribution of the six tidal interaction models (T1, T2, T3, T4, T5, and T6)
with different orbital configurations and the isolated  disc model D1 on
$E_{\rm sf}$-$E_{\rm gc}$ plane (left) 
and on $S_{\rm N,R(global)}$-$S_{\rm N(global)}$ plane (right).
D1, T1 (fiducial), T2 (retrograde-retrograde), 
T4 (LSB), T5 (lower disc mass), T6 (very low disc mass)
and T3 (highly inclined discs),
are  represented by
open triangle, open square, open pentagon, open circle,
filled triangle, filled square, and filled pentagon, respectively. 
}
\label{Figure. 19}
\end{figure*}

\begin{figure*}
\psfig{file=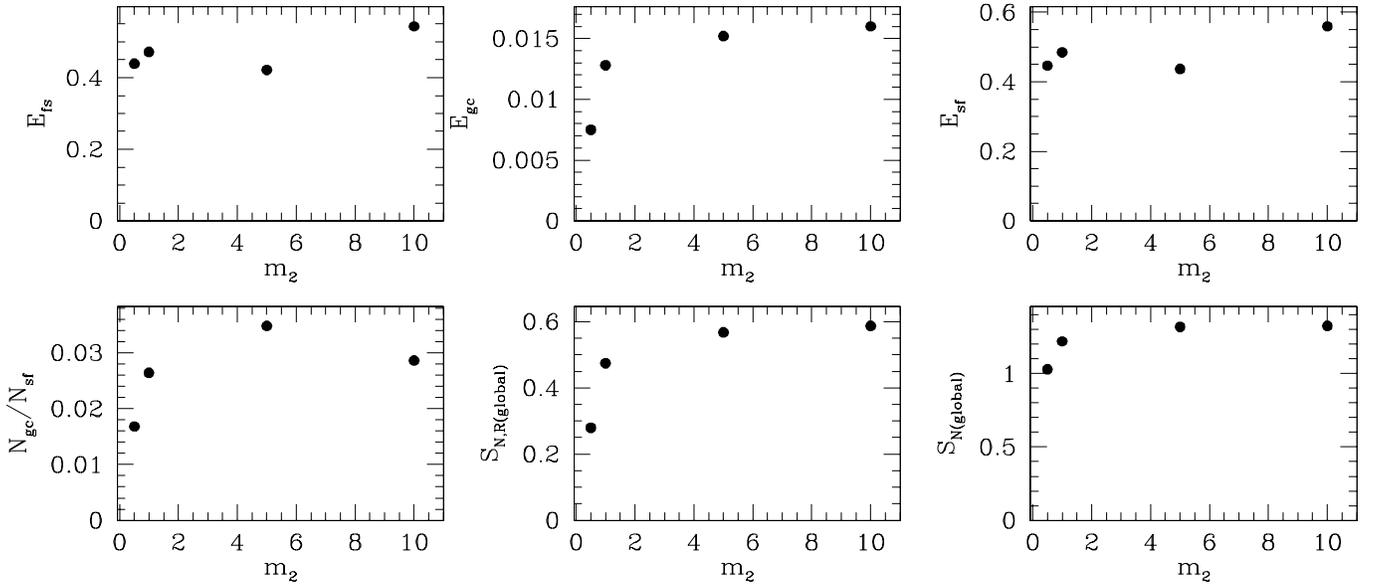,width=18.cm}
\caption{ 
The dependence of  $E_{\rm fs}$ (upper left),
$E_{\rm gc}$ (upper middle), $E_{\rm sf}$ (upper right),
$N_{\rm gc}/N_{\rm sf}$ (lower left),   $S_{\rm N,R(global)}$ (lower middle),
and $S_{\rm N(global)}$ (lower right) on the ratio of two interacting  discs
(represented by $m_{2}$) for the tidal interaction  models T1, T7, T8, and T9.
Physical meaning of the above six quantities are given in the text. 
}
\label{Figure. 20}
\end{figure*}

\subsection{Tidal interactions}

\subsubsection{Difference in globular cluster  properties between tidal interaction and merging}

Formation processes of the MRC 
are found to be essentially the same between the merger  and tidal interaction
models  in the present study. 
Here we focus on the following two remarkable differences in physical properties
of the MPC and the MRC between the merger model M1  and the tidal interaction one T1.
Firstly, although strong tidal forces due to galaxy interaction
can compress the interstellar gas and  dramatically increase  the gas pressure
to the threshold value of molecular collapse,
the formation efficiency of the MRC ($E_{\rm gc}$) is smaller (0.0128) 
in an interaction than  
a  merger ($E_{\rm gc}$ = 0.0298). 
This is firstly because the tidal force, the strength of which can be measured roughly by
${r_{\rm p}}^{-2}$, is significantly weaker  in  T1 with $r_{\rm p}$ = 2.0
than in  M1 with  $r_{\rm p}$ = 0.5,
and secondly because hydrodynamical interaction between two gas discs,
which also greatly contributes to the formation of shocked gaseous regions
with high density and pressure, does  not occur in an interaction.   
As a natural result of low $E_{\rm gc}$,
T1 shows lower $S_{\rm N(global)}$ (1.22) than 
M1 (1.78).
Other quantities such as $E_{\rm fs}$, $E_{\rm sf}$, and 
$N_{\rm gc}/N_{\rm sf}$ are all smaller in T1 than in M1 (compare those of T1
with those of  M1 in Table 1).

Secondly, as is shown in Fig. 17,
the final morphology of the disc in T1 appears to be similar
to an S0 rather than to a spiral owing to the less pronounced 
spiral arms and the dynamically thickened disc.
Fig. 18 furthermore shows that although the MPC keep their initial spherical distribution,
the MRC show a highly flattened distribution after the formation of the MRC because of 
tidal interaction. The spatial distribution of the MRC in T1 is reminiscent of
the Galactic thick disc component.
Since dynamical relaxation and  tidal stripping of the MPC can  not so severely
affect the spatial distribution of the MPC  in T1 compared with M1,
the half mass radius of the MPC and the radius within which 90 \% of the MPC 
are included are not significantly  changed from the initial values.
Furthermore, owing to the disc-like distribution of the MRC,
the half-mass radius of the MRC in T1 is much larger (0.35) than that in M1 (0.05).
These results suggest that if S0s are formed from tidal interaction between disc galaxies,
they should have (1) thick stellar discs composed of old stars, (2) moderately
high $S_{\rm N(global)}$ (higher than those of late-type spirals but lower than those
of ellipticals), (3)  MRC with highly flattened spatial distributions,
and (4) larger  half-mass radius  for  the MRC. 
The origin of S0s is discussed later in \S 4 in terms of the observed physical properties
of the MPC and the MRC of S0s.

\subsubsection{Dependence on initial conditions}

The dependences of 
structure, kinematics, and chemical properties
of the MPC and the MRC  on initial condition of tidal interaction
are found to be rather complicated compared with the merger models
in the present study.
Figures  19 and 20  summarise the important dependences on
orbital configurations and masses of tidally interacting galaxies 
and on the mass-ratio of these, respectively.  
We describe these as follows:

 (i) The range in orbital configurations of tidal interaction introduces
the diversity in $E_{\rm gc}$, $E_{\rm sf}$,
$S_{\rm N(global)}$, and $S_{\rm N, R(global)}$ (see Fig. 19).
This is basically similar to the derived results of the merger models.
There are correlations between these quantities:
$E_{\rm gc}$ is proportional to $E_{\rm sf}$ whereas $S_{\rm N(global)}$
is proportional to $S_{\rm N,R(global)}$.

 (ii) Irrespective of parameters in equal-mass interaction models,
$S_{\rm N(global)}$ in these models is systematically smaller  
than that in the major (equal-mass) merger models
(compare  Figures 14  and 19). 
This is essentially because
tidal forces are  weaker in the tidal interaction model than in the merger models.

 (iii) Unlike the merger models, tidal interaction models 
 show appreciably smaller $E_{\rm gc}$ in the retrograde-retrograde interaction case (0.0082)
than in the prograde-prograde one (0.0128). This is because a tidal stellar bar
which induces strong tidal shocks in gaseous regions can not be developed
in the retrograde-retrograde interaction model.

 (iv) The total number of the MRC formed 
in tidal interaction between low surface-brightness galaxies (LSB)
is rather small  ($E_{\rm gc}$ = 0.0037).
This suggests that the formation efficiency of the MRC in tidally interacting 
disc galaxies strongly depends
on the surface-brightness of the discs.

 (v) Lower mass galaxies are likely to show a lower formation efficiency of the MRC 
(e.g., $E_{\rm gc}$ = 0.0 in the T6 model with $m_{\rm d}$ = 0.01),
even if these have the same central surface-brightness as that of higher mass 
galaxies.
The above results (iv) and the observational fact (Impey \& Bothun 1997)
that low luminosity spirals  are likely to have low surface-brightnesses
contributes to  this tendency for  low mass galaxies. 
The essential reason for this is that although the strength of tidal
force becomes significantly weaker as the masses of interacting galaxies
becomes smaller (because the strength of tidal force at the pericentre distance
of an interacting galaxy depends roughly on the mass
owing to the adopted mass-size relation of disc galaxies),
the threshold pressure for molecular collapse remains the same.
The same trend can be seen in the low mass LSB merger models (see the results of M15).
These results suggest that the formation of the MRC is likely to be  inefficient
in very low mass interacting and merging galaxies. 

(vi)  $E_{\rm gc}$, 
$S_{\rm N,R(global)}$,  and $S_{\rm N(global)}$ are all likely to be larger
for the  model with the  larger $m_{2}$.  
These dependences are nearly the same as those derived in the merger models.

(vii) High speed tidal encounters  between a small disc (with $m_{d}$ = 0.1) 
and a large disc (with $m_{d}$ = 1.0 and $m_{2}$ = 10),
which correspond to the so-called ``galaxy harassment'' concept (Moore et al. 1996),
can trigger the formation of
the MRC with a  formation efficiency of the MRC similar to those of other models
(see the results of T10 and T11 in Table 1 and compare these results
with those of other tidal interaction models).

\section{Discussion}


\subsection{$S_{\rm N}$ problems in elliptical galaxies}

The observed difference in the specific frequency $S_{\rm N}$ of
globular clusters between spiral galaxies and ellipticals had
been used to argue against the major merger scenario 
of elliptical galaxy formation (e.g., van den Bergh 1990;
Harris 1991).  These arguments are based on the assumption that
dissipationless major merging can not increase the total number
of globular clusters per unit starlight.
Schweizer (1987) suggested that new globular clusters can be formed
in gas-rich major mergers owing to the strong tidal shocks
and thus may increase $S_{\rm N}$ of the globular cluster system around 
merger remnant galaxies.
Ashman \& Zepf (1992) and Zepf \& Ashman (1993) suggested
that the increase of $S_{\rm N}$ is due to metal-rich globular cluster
formation which results in  the bimodal colour distribution
of globular clusters around elliptical galaxies.
Here we have  demonstrated that 
gas-rich major mergers can dramatically increase 
the formation efficiency of globular clusters  
due to strong tidal shocks in gaseous regions, increasing 
the $S_{\rm N}$  in elliptical galaxies formed by merging. 
Thus our study clearly supports these previous claims 
by Schweizer (1987), Ashman \& Zepf (1992), and Zepf \& Ashman (1993)
that $S_{\rm N}$ can increase in dissipative merging,
and hence offers a possible explanation as to 
why $S_{\rm N}$ is higher in elliptical galaxies than in spirals.

However, as pointed out by several authors 
(e.g., Forbes et al. 1997; Kissler-Patig et al. 1999; C\^ote et
al. 1998), the $S_{\rm N}$ values for elliptical galaxies 
are {\it not} adequately explained by the merger scenario. For example,
Forbes et al. (1997) found that the high $S_{\rm N}$ ($\geq$ 5)
observed in cluster giant ellipticals is associated
with a larger number of  metal-poor clusters, not metal-rich
ones. Our numerical study has demonstrated that in order to obtain
high $S_{\rm N}$ in ellipticals, we must produce a larger number
of metal-rich globular clusters, which is obviously inconsistent 
with the observational results of Forbes et al. (1997).
So, although the merger scenario 
can explain the origin of globular 
cluster systems with low $S_{\rm N}$ values, i.e., $\le$  3,
it still appears to be unable to  explain  the 
high $S_{\rm N}$ systems. An 
alternative  scenario  may be required to explain of the
origin of the metal-poor globular clusters in high $S_{\rm N}$
elliptical galaxies. 
One caveat to this is that we have assumed an $S_{\rm N}$ value for the
progenitor discs of 0.8. Although this is a fairly typical value
for a spiral galaxy, they reveal a wide range from $\sim$0.3 to
$\sim$3 (Ashman \& Zepf 1998). The merger of high $S_{\rm N}$
spirals would help to explain the potential $S_{\rm N}$ problem
highlighted above. 


Our present study has raised another problem related to $S_{\rm
N}$ that has not been realized in previous studies: The local
$S_{\rm N}$ decreases with radius from the centre of an
elliptical galaxy formed by major merging, which is inconsistent
with observations.  The essential reason for this radial
dependence is that the MRC form a centrally concentrated
distribution compared with the host merger's field star
distribution (this is related to the merger models overproducing
the number of MRC).  
One way to avoid this difficulty is to assume that the
newly formed globular clusters formed in the central region can
be preferentially destroyed by the very strong tidal field of the
central region of the merger.  If the central globular clusters
can be destroyed more preferentially by some physical processes
such as bulge and disc shocking (e.g 
Aguilar, Hut, \& Ostriker 1988 and 
Gnedin \& Ostriker 1997), then the resultant number
surface density distribution may well be shallower than that of
field stars so that the local $S_{\rm N}$ can increase with
radius. There are however no detailed numerical studies to
determine whether this ``selective'' tidal destruction of central globular
clusters can really happen in {\it the time-dependent
gravitational fields of major mergers}. A future dynamical
study should investigate the tidal effects in the central region
of a merger on the structural evolution of the MRC in order to
determine whether the observed radial dependence of $S_{\rm N}$
is an additional problem for the merger model of globular
cluster formation.

\subsection{The origin of bimodal colour distributions of globular clusters}

Most giant elliptical galaxies are observed to have bimodal
colour distributions of globular clusters which most likely
result from two sub-populations with distinct metallicities and
ages (e.g., Larsen et al. 2001; Kundu \& Whitmore 2001).  Ashman
\& Zepf (1992) suggested that if elliptical galaxies are
formed by dissipative and star-forming mergers between spirals,
they should reveal a subpopulation of 
metal-poor globular clusters (those initially located
in the progenitor spirals) and metal-rich ones (formed in the
chemically-enriched gas of the spirals). 

Our study confirms that dissipative and star-forming merging
can create new globular clusters, however it can not correctly 
reproduce the bimodal metallicity distributions 
seen in  elliptical galaxies.
We have shown in \S 3
that the mean metallicity of the MRC in an elliptical galaxy
formed by {\it recent major merging} is more than twice solar (or
[Fe/H] $\ge$ +0.3). 
This result holds even if we 
change the input parameters, i.e., 
the instantaneous chemical mixing approximation
and the adopted negative metallicity gradients in discs.
This metallicity is significantly greater than the 
typical value of [Fe/H] $\sim$ $-0.5$ observed photometrically
and spectroscopically for many ellipticals (e.g. Forbes 2001).

A possible explanation for this difference is that the 
present study significantly overestimates 
the metallicities of the MRC.
If MRC are developed preferentially in the metal-poor outer disk regions
(not in the metal-rich central ones at all) and formed from gas clouds
only in the earlier merging epoch when chemical evolution 
has not yet increased the metallicity of gas clouds,
then the mean metallicity of MRC can be significantly lowered
and the above problem can be avoided.
Since there is no observational evidence for such biased
formation of globular clusters in mergers,
this interpretation seems unlikely. 
Our results imply  that recent major mergers do not result in 
elliptical galaxies with the observed globular cluster 
metallicity distributions 
and therefore the merging epoch must have been at sufficiently
early epcohs so that the progenitor discs had 
low gaseous metallicities ([Fe/H] $<$ $-0.5$).

\subsection{Correlations with the Hubble morphological sequence}



The mean $S_{\rm N}$ for S0s is observed to be 1.0$\pm$0.6 (Kundu
\& Whitmore 2001) which is higher than that of late type spiral
galaxies ($S_{\rm N}$ = 0.5$\pm$0.2 in Harris 1991) and lower
than that of elliptical ones ($S_{\rm N}$ = 2.4$\pm$1.8 in Kundu
\& Whitmore 2001). The observed $S_{\rm N}$ for S0s suggests that
for a late type spiral to become an S0, $S_{\rm N}$ needs to
increase by some mechanism and thus S0s can not be transformed
from late-type spirals via mechanisms which are highly unlikely
to form new star clusters in discs (e.g., ram pressure
stripping).  Minor merging and unequal-mass mergers have been
demonstrated to transform two discs into an S0 owing to the
strong dynamical heating and the triggered central star-bursts
that grow in the bulge (e.g., Barnes 1996;
Bekki 1998).  Tidal interaction of galaxies has also been suggested 
as being responsible for S0 formation (Noguchi 1988; Bekki et
al. 2001).  Our study has confirmed these earlier numerical
results and demonstrated that $S_{\rm N}$ can be increased by a
factor of 1.5 during star-bursts triggered by unequal-mass or
minor merging and tidal interaction, consistent with the observed
$S_{\rm N}$ avlues for S0s.  These results suggest that the observed
$S_{\rm N}$ in S0s can be understood in terms of the
strength of gravitational interaction that controls both galaxy
morphological types and the formation of new globular clusters 
(and thus $S_{\rm N}$).

We suggest that the origin of $S_{\rm N}$ in S0s is
closely associated with the formation of new globular clusters in
unequal-mass merging and interacting galaxies. In particular, it 
depends on (1)
$S_{\rm N}$ for the very metal-rich globular clusters 
, (2) the age of these
metal-rich globular clusters, and (3) their density distribution and
kinematics.  The present numerical results show that 
$S_{\rm N}$ increases during the morphological transformation
(from a late-type spiral into an early-type S0) due to the
formation of new very metal-rich globular clusters 
and that the new globular cluster system has a
more flattened distribution.  Although Kundu \& Whitmore (2001)
revealed that the mean metallicity of the globular cluster system of S0s
is primarily a function of the host galaxy's luminosity 
(or mass), it is not clear from their results whether the $S_{\rm
N}$ for very metal-rich globular clusters in S0s is higher than
that of spirals.  The ages and kinematics of MRC in S0s have not
been yet determined by spectroscopic
observations.

\section{Conclusions}

We have numerically investigated how the global dynamical
evolution of merging and interacting galaxies can determine the
formation processes and fundamental physical properties of
globular clusters.  We have assumed that a globular cluster can
be formed from a giant molecular cloud in a galaxy if the warm
interstellar gas of the galaxy becomes high enough ($P_{\rm gas}$
$>$ $10^5$ $k_{\rm B}$).  We summarise our principle results as
follows.

(1) Strong tidal shocks induced by galaxy merging and interaction
can dramatically increase gaseous pressure ($P_{\rm g}$ $>$
$10^5$ $k_{\rm B}$) so that molecular clouds can collapse to form
globular clusters.
During the formation of globular clusters in a
merging/interacting galaxy, the ratio of the formation rate of
globular clusters to that of field stars increases due to the
larger fraction of gas with high pressure in the galaxy.  
Thus globular cluster formation is more 
efficient in star-burst regions of galaxies.
This result can explain why young globular clusters are commonly observed
in merging and interacting galaxies compared to 
isolated spirals.

(2) The specific frequency ($S_{\rm N}$) of globular cluster
systems can
be increased a factor of $2-3$ in a major gaseous merger (which
results in the formation of an elliptical galaxy) 
due to the creation of new metal-rich globular clusters.  
However, many 
elliptical galaxies are observed 
to have higher $S_{\rm N}$ values and 
higher ratios of metal-poor to metal-rich clusters than can
be explained by our merger simulations. 
Merger progenitor spirals with $S_{\rm N}$ values higher than
those assumed here (i.e., 0.8) may help aleviate this problem.



(3) The metallicity distribution of metal-rich globular cluster
systems formed by major merging depends upon the initial
metallicity distribution of merger progenitor discs and the
chemical evolution of gas. In present-day mergers, 
the mean value of the metallicity distribution of newly
formed globular clusters is typically higher than twice solar
metallicity because of the initially large gas metallicity 
and efficient chemical processing. 
This means that recent major mergers can yield
bimodal globular cluster colour distributions but, these
distributions have metal-rich mean values that are much greater
than that observed in elliptical
galaxies(i.e, [Fe/H] $\sim$
$-0.5$). This suggests that if most of elliptical galaxies are
formed by major merging, then the typical merging epoch should be a
high redshift when the merger progenitor discs have
low-metallicity gas.

(4) The dynamical evolution of metal-poor globular clusters
(initially in the halo regions of merger progenitor spirals) is
different from that of the newly formed metal-rich clusters.  
Metal-poor globular clusters experience stronger 
angular momentum transfer from the inner to outer regions, 
whereas metal-rich clusters experience a larger amount of gas
dissipation prior to their formation.  As a natural result of
this, the surface density distribution and kinematical
properties of the two globular cluster subpopulations 
are different. 
For example, the metal-poor globular clusters
show shallower density profiles, larger velocity dispersion
in the central regions and a large amount of rotation in the
outer regions.

(5) The $S_{\rm N}$, structural, and kinematical properties of
globular clusters in the remnants of major mergers depend weakly
on the orbital configurations of the merger (such as initial
orbital angular momentum and whether a merger is 
prograde-prograde or retrograde-retrograde).  The metallicity
distribution of the newly formed globular cluster system does not
depend on the above orbital parameters.
Given the observed positive correlation between luminosity
and gas metallicity in disc galaxies (Zaritsky et al. 1994),
our results suggest that there should be a positive correlation
between metallicities of metal-rich globular clusters and
luminosities of their host galaxies.

     \indent (6) The formation efficiency, total number, and $S_{\rm
     N}$ of globular clusters formed in mergers depends on the
     mass ratio of the two merging discs ($m_{2}$) in such a way that
     each of these three quantities are smaller in mergers with
     smaller $m_{2}$. This is due to the weaker galactic
     tide in mergers with smaller $m_{2}$, as only a smaller
     fraction of gas can have enough high gas pressure
     ($P_{\rm g}$ $>$ $10^5$ $k_{\rm B}$) to trigger the
     molecular cloud collapse leading to globular cluster
     formation. These results imply that S0s, which can be formed
     by minor and unequal-mass merging, will show $S_{\rm N}$
values that are 
     appreciably higher than that of typical spirals, but 
     lower than that of ellipticals.  These results further
     suggest that spiral galaxies with thick disc components,
     which can be formed by minor merging, should also show $S_{\rm
     N}$ values slightly higher than that of spirals without thick
     discs.\\ 

\indent (7) The $S_{\rm N}$ of globular clusters is
     more likely to be lower in tidally interacting galaxies than
     in merging ones.  The formation efficiency, total number, and
     $S_{\rm N}$ of globular clusters in interacting galaxies
     strongly depends on the structure of disc galaxies, orbital
     configurations, and the mass ratio of two
     interacting discs.  For example, $S_{\rm N}$ does not
     increase significantly in low surface-brightness galaxies
     compared to high surface-brightness galaxies.
     Furthermore $S_{\rm N}$ is higher in an interacting galaxy
     with a larger mass-ratio.  
As tidal interaction
     may also transform gas-rich spirals into a gas-poor S0s, our
     tidal interaction models can provide a natural explanation
     for the observed $S_{\rm N}$ values of S0s.\\

\indent (8) The formation efficiency of globular clusters in 
merging and interacting
galaxies is likely to decrease with galaxy mass, though field
star formation is still ongoing in those galaxies.  This implies
that there could be a minimum galactic mass for a galaxy to
harbour globular clusters.  This result also provides an
important implication for the formation of metal-poor halo
globular clusters that are believed to be formed more than 10 Gyr
ago, if such old globular clusters are formed from merging 
of less massive sub-galactic gaseous clumps at
high redshift.

\section{Acknowledgement}
We are  grateful to the anonymous referee for valuable comments,
which contribute to improve the present paper.
KB and WJC  acknowledge the financial support of the Australian Research Council
throughout the course of this work.
MB would like to thank the Royal Society for its fellowship grant.
We thank  R. Balasubramanyam  for useful discussions about star formation.



\begin{table*}
\centering
\begin{minipage}{185mm}
\caption{Model parameters and results of galaxy merging and interaction.}
\begin{tabular}{cccccccccccccl}
Model no. & 
$m_{\rm d}$  
($\times$ $M_{\rm d}$)  
& $m_{2}$ 
& $r_{\rm p}$ ($\times$ $r_{\rm d}$)
&orbit  
&$C_{\rm gc}$ 
&{$E_{\rm sf}$%
\footnote{Mass ratio of newly formed stellar components (field stars and MRC) to initial gas.}}
& {$E_{\rm fs}$%
\footnote{Mass ratio of newly formed field stars to initial gas.}} 
& {$E_{\rm gc}$%
\footnote{Mass ratio of MRC  to initial gas.}}  
& {$N_{\rm gc}/N_{\rm sf}$%
\footnote{Number  ratio of MRC to newly formed stellar components for initial $S_{\rm N}$
of spirals = 0.8.}} 
& {$S_{\rm N}$%
\footnote{Specific frequency of globular clusters 
for initial $S_{\rm N}$ of spirals = 0.8.}}
& {$N_{\rm MPC}/N_{\rm MRC}$%
\footnote{Number ratio of MPC to MRC for initial $S_{\rm N}$ of spirals = 0.8.}}
& Comments \\
D1 & 1.0 &  --  & --  &  -- & 0.1 & 0.515 & 0.509 & 0.0059 & 0.011 & 0.96 & 3.39&isolated  \\
M1 & 1.0 &  1.0 & 0.5 &  PP & 0.1 & 0.756 & 0.726 & 0.0298 & 0.039 & 1.78 & 0.67&fiducial  \\
M2 & 1.0 &  1.0 & 0.5 &  PR & 0.1 & 0.848 & 0.819 & 0.0289 & 0.034 & 1.73 & 0.69&\\
M3 & 1.0 &  1.0 & 0.5 &  RR & 0.1 & 0.917 & 0.868 & 0.0485 & 0.053 & 2.40 & 0.41&\\
M4 & 1.0 &  1.0 & 0.5 &  HI & 0.1 & 0.752 & 0.723 & 0.0229 & 0.031 & 1.54 & 0.87&\\
M5 & 1.0 &  1.0 & 1.0 &  LA & 0.1 & 0.760 & 0.733 & 0.0276 & 0.036 & 1.70 & 0.72&\\
M6 & 1.0 &  1.0 & 0.5 &  PP & 0.1 & 0.813 & 0.779 & 0.0344 & 0.042 & 1.88 & 0.58&  $f_{\rm g}$=0.2\\
M7 & 1.0 &  1.0 & 0.5 &  AN & 0.1 & 0.602 & 0.581 & 0.0212 & 0.035 & 1.50 & 0.95& Antennae \\
M8 & 1.0 &  1.0 & 0.5 &  AN & 1.0 & 0.629 & 0.476 & 0.1530 & 0.242 & 6.41 & 0.13&Antennae \\
M9 & 1.0 &  1.0 & 0.5 &  PP & 0.5 & 0.756 & 0.549 & 0.2075 & 0.275 & 8.35 & 0.10&\\
M10& 1.0 &  1.0 & 0.5 &  PP & 1.0 & 0.759 & 0.350 & 0.4090 & 0.539 & 16.2 & 0.05&\\
M11& 1.0 &  0.1 & 0.5 &  PP & 0.1 & 0.586 & 0.573 & 0.0128 & 0.022 & 1.15 & 1.56&minor merger\\
M12& 1.0 &  0.3 & 0.5 &  PP & 0.1 & 0.748 & 0.721 & 0.0268  & 0.036 &1.50 & 0.75& unequal-mass\\
M13& 1.0 &  0.5 & 0.5 &  PP & 0.1 & 0.769 & 0.740 & 0.0291 & 0.038 & 1.50 & 0.69&\\
M14& 0.01&  1.0 & 0.5 &  PP & 0.1 & 0.807 & 0.765 & 0.0420 & 0.052 & 2.20 & 0.48&\\
M15& 0.01&  1.0 & 0.5 &  PP & 0.1 & 0.262 & 0.260 & 0.0020 & 0.009 & 0.85 & 8.89&LSB\\
T1 & 1.0 &  1.0 & 2.0 &  PP & 0.1 & 0.485 & 0.472 & 0.0128 & 0.026 & 1.22 & 1.57& interaction\\
T2 & 1.0 &  1.0 & 2.0 &  RR & 0.1 & 0.331 & 0.323 & 0.0082 & 0.025 & 1.07 & 2.44&\\
T3 & 1.0 &  1.0 & 2.0 &  HI & 0.1 & 0.446 & 0.437 & 0.0089 & 0.020 & 1.08 & 2.26&\\
T4 & 1.0 &  1.0 & 2.0 &  PP & 0.1 & 0.123 & 0.119 & 0.0037 & 0.031 & 0.93 & 5.33&LSB\\
T5 & 0.1 &  1.0 & 2.0 &  PP & 0.1 & 0.710 & 0.698 & 0.0119 & 0.017 & 1.15 & 1.69&\\
T6 & 0.01&  1.0 & 2.0 &  PP & 0.1 & 0.001 & 0.001 & 0.0000 & 0.000 & 0.80 & -&\\
T7 & 1.0 &  0.5 & 2.0 &  PP & 0.1 & 0.446 & 0.439 & 0.0075 & 0.017 & 1.03 & 2.68&\\
T8 & 1.0 &  5.0 & 2.0 &  PP & 0.1 & 0.437 & 0.422 & 0.0152 & 0.035 & 1.32 & 1.32&\\
T9 & 1.0 &  10.0& 2.0 &  PP & 0.1 & 0.559 & 0.543 & 0.0160 & 0.029 & 1.33 & 1.26&\\
T10& 0.1 &  10.0& 2.0 &  PP & 0.1 & 0.663 & 0.635 & 0.0284 & 0.043 & 1.76 & 0.71&\\
T11& 0.1 &  10.0& 2.0 &  PP & 0.1 & 0.725 & 0.706 & 0.0193 & 0.027 & 1.41 & 1.04& $e_{\rm p}$ = 5 \\

\end{tabular}
\end{minipage}
\end{table*}


\newpage





\end{document}